\documentclass[onecolumn]{revtex4-2}

\usepackage{amsmath}
\usepackage{amsfonts}
\usepackage{amssymb}
\usepackage{graphicx}
\usepackage{hyperref}

\begin{document}


\title{Quantization of a New Canonical, Covariant, and Symplectic Hamiltonian Density}

\author{David Chester}
\email{DavidC@QuantumGravityResearch.org}
\affiliation{Quantum Gravity Research, Topanga, CA, USA}

\author{Xerxes D.~Arsiwalla}
\email{x.d.arsiwalla@gmail.com}
\affiliation{Pompeu Fabra University, Barcelona, Spain}
\affiliation{Wolfram Research, Champaign, IL, USA}

\author{Louis H.~Kauffman} 
\email{loukau@gmail.com}
\affiliation{University of Illinois at Chicago, Chicago, IL, USA}

\author{Michel Planat}
\email{michel.planat@femto-st.fr}
\affiliation{CNRS, Institut FEMTO-ST, Université de Franche-Comté, F-25044 Besançon, France}

\author{Klee Irwin}
\email{Klee@QuantumGravityResearch.org}
\affiliation{\, Quantum Gravity Research, Topanga, CA, USA \,}

\date{\today}

\begin{abstract}
We generalize Koopman-von Neumann classical mechanics to poly-symplectic fields and recover De Donder-Weyl theory. 
Comparing with Dirac's Hamiltonian density inspires a new Hamiltonian formulation with a canonical momentum field that is Lorentz covariant with symplectic geometry. 
We provide commutation relations for the classical and quantum fields that generalize the Koopman-von Neumann and Heisenberg algebras. 
The classical algebra requires four fields that generalize space-time, energy-momentum, frequency-wavenumber, and the Fourier conjugate of energy-momentum. 
We clarify how 1st and 2nd quantization can be found by simply mapping between operators in classical and quantum commutator algebras.
\end{abstract}


\maketitle

\setcounter{tocdepth}{2}

\tableofcontents


\section{Introduction}



Koopman-von Neumann (KvN) mechanics formulates classical mechanics (CM) with a complex wavefunction in a Hilbert space \cite{Koopman,vonNeumann,Bondar,Morgan:2019azd,Morgan:2021jeh,Piasecki:2021keq,opmech2022}. This formulation helps clarify the similarities and differences between CM and quantum mechanics (QM). Bondar et al.~studied an algebra for KvN mechanics that can be quantized by mapping to the Heisenberg algebra \cite{Bondar}. The Koopman-von Neumann algebra contains the position operator $\hat{x}^i$, its Fourier-conjugate wavenumber $\hat{k}_j$, the momentum operator $\hat{p}_k$, and its Fourier conjugate $\hat{q}^l$. By recognizing that this KvN algebra has Fourier conjugate variables over phase space, quantization can be found by setting $\hat{p}_i = \hbar \hat{k}_i$. Our primary goal is to generalize this KvN quantization to relativistic field theories. 

De Donder-Weyl (DDW) theory contains a covariant Hamiltonian density for relativistic field theories with DDW equations \cite{DeDonder1935,Weyl1934,Weyl1935,vonRieth}. 
DDW theory contains poly-symplectic geometry, which introduces a conjugate poly-momentum field of a higher tensor rank than the field; the prefix poly refers to the higher rank. In classical mechanics, the Euler-Lagrange equations account for a time derivative of the momentum. In relativistic field theory, the analogous conjugate momentum is the poly-momentum defined with a partial derivative, as found in DDW theory. DDW theory can be contrasted with Dirac's canonical Hamiltonian density \cite{Dirac1950}, which uses a partial time derivative for the canonical momentum despite the Euler-Lagrange equations containing the partial derivative.

Our initial goal was to recover KvN quantization of DDW theory by generalizing KvN mechanics to poly-symplectic fields. We find that the DDW equations can be found from a poly-Liouville operator, which is related to previously studied poly-symplectic Poisson brackets \cite{Kanatchikov:1999ut}, but this obscures Dirac's canonical quantization.  However, new poly-KvN commutator algebras allow for straightforward quantization by mapping commutators of classical operators to quantum operators. While this provides a new path towards DDW quantization, Kanatchikov has extensively discussed precanonical quantization as the (geometric) quantization of DDW theory by considering Gerstenhaber brackets as generalized Poisson brackets \cite{Kanatchikov:1996fx, Kanatchikov:1997pj, Kanatchikov:1997wp, Kanatchikov:1998yu, Kanatchikov:1998xz, Kanatchikov:1998vt, Kanatchikov:1999ut, Kanatchikov:2001uf, Kanatchikov:2002yh,Kanatchikov:2008he,Kanatchikov:2012gr, Kanatchikov:2013xmu, Kanatchikov:2015hna, Kanatchikov:2017zfe, Kanatchikov:2018uoy}.

\begin{figure}
  \centering
  \includegraphics{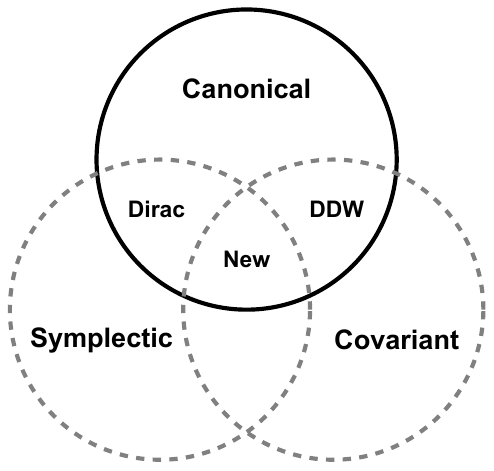}
  \caption{A comparison of the Dirac $\mathcal{H}_D$, De Donder-Weyl $\mathcal{H}$, and the new covariant $\mathcal{H}_{new}$ Hamiltonian densities, focusing on whether these formulations are covariant or symplectic. All Hamiltonian formulations contain a canonical momentum, but not all are covariant or symplectic. These three Hamiltonian formulations arise from the same Lagrangian density $\mathcal{L}$. }
  \label{fig:ccs}
\end{figure}

The main result of this work is to find the KvN quantization of a new Hamiltonian density that is canonical, covariant, and symplectic. Canonical quantization is typically used with Dirac's canonical Hamiltonian density, which is symplectic but not covariant \cite{Dirac1950}. Precanonical quantization can be applied to the DDW theory \cite{Kanatchikov:1998vt,Kanatchikov:2001uf}, whose Hamiltonian density is covariant but not symplectic. Comparing these two formulations inspires a new covariant, canonical, and symplectic Hamiltonian density, as shown in Fig.~\eqref{fig:ccs}. We introduce a generalized KvN algebra for these fields, whose 2nd quantization leads to canonical commutation relations in terms of a covariant phase space of fields with symplectic geometry.

While Witten and Crnkovic had proven that covariant and symplectic structure for fields exists, this focused on identifying a symplectic charge that integrates over a hypersurface with time-like boundary \cite{Witten1986, Zuckerman, Witten1987, Crnkovic1987, Crnkovic1988}, rather than considering a new type of Hamiltonian density found from the canonical structure of the fields. 
Covariant formulations of Hamiltonian dynamics have been previously discussed, but often do not present a Hamiltonian formulation, despite referring to the Hamiltonian dynamics \cite{Frauendiener, Harlow:2019yfa}. In hindsight, Iyer and Wald's formulation is close to ours, as a time-like vector $t^a$ was introduced in a similar manner to our $\hat{\tau}^\mu$ to find their Hamiltonian $H_X$ to study energy; however, no Hamiltonian density was found directly from the Lagrangian density in this formulation for gravitational fields \cite{Iyer:1994ys,Iyer:1995kg,Wald:1999wa}. Ashtekar found a symplectic Hamiltonian density that used a hypersurface based on null infinity, but this was restricted to the study of general relativity \cite{Ashtekar}. Covariant phase space methods have been recently explored in a wide class of gravitational theories \cite{Margalef-Bentabol:2020teu,G:2021xvv,BarberoG:2021cei,G:2022ger}. However, a comprehensive formulation of arbitrary field theory with a covariant, symplectic, and canonical Hamiltonian density still appears to be lacking to the best of our knowldge.


To the best of our knowledge, this work includes the following results for the first time (or at least provides additional context and clarity): 
\begin{itemize}
\item Construction of proper relativistic KvN mechanics and relativistic KvN algebra.
\item Generalization of KvN mechanics to poly-KvN fields as DDW theory with a new poly-KvN algebra. 
\item A new covariant and symplectic Hamiltonian density formulation for relativistic fields.
\item 2nd quantization via deformation of commutator algebras over fields.
\end{itemize}


This manuscript is organized as follows. Section \eqref{quant-new-H} introduces the new Hamiltonian density, the analogous set of Hamilton's equations, the generalized KvN algebra in terms of a Fourier-phase space of fields, and its KvN quantization. Appendix \eqref{KvN} focuses on a relativistic formulation of KvN mechanics and its quantization to give the relativistic generalization of the Heisenberg algebra. Appendix \eqref{DDW} introduces DDW theory, demonstrates that the poly-symplectic generalization of KvN mechanics is equivalent to DDW theory, and presents KvN quantization of DDW theory. Readers unfamiliar with KvN or DDW theory may prefer to start with the appropriate appendices, while experts may proceed directly to Section ~\eqref{quant-new-H}. While the appendices may contain some new results, they often contain rederivations of older results. 



\section{Quantization of a new Hamiltonian density}
\label{quant-new-H}

In this section, a new Hamiltonian density inspired by Dirac's canonical Hamiltonian density and the De Donder-Weyl Hamiltonian density. Our strategy is to develop a new Hamiltonian formulation by taking the local time-like component of a frame field combining it with the DDW canonical momentum to obtain something more similar to the Dirac canonical momentum. By contracting a time-like unit vector $\hat{\tau}_{\mu}(x) = e_\mu^0(x)$ with the poly-momentum field to give $\pi(x) = \hat{\tau}_\mu \pi^\mu(x)$, the new Hamiltonian density limits to Dirac's Hamiltonian density when the unit vector is $\hat{\tau}^{\mu}(x)=(1,0,0,0)$. In this manner, we can locally refer to different spacetime foliations with applications for dynamical foliation schemes. The new canonical momentum field is found to be covariant with respect to the global manifold, yet is symplectic with respect to the (position) field $\phi(x)$. This new approach is well suited for canonical quantum relativistic evolution with manifestly covariant and symplectic fields.


A key result is the canonical and covariant commutation relations of a scalar field $\phi$ with its conjugate field for classical, 1st, and 2nd quantized fields. For classical fields $\phi(x,p_x)$ over phase space (inspired by KvN mechanics), a Fourier-conjugate wavenumber field $\kappa(y,p_y)$ is found, which can be first quantized to give $\phi(x)$ and $\kappa(y)$ as classical fields over spacetime instead of phase space. Table~\eqref{table:012} summarizes the different possible phase space configurations of classical and quantum fields. Note that $\phi(x)$ is typically referred to as a classical field, but in our context, first quantization of KvN mechanics over fields implies that classical fields such as $\phi(x)$ are actually first quantized, which is expanded upon in Section \eqref{classical-fields}. Quantum fields are depicted by capital letters, which lead to $\Phi(x)$ and the conjugate momentum $\Pi(y)$ as second quantized fields, 
\begin{eqnarray}
&\mbox{0th quantized (classical): }& [\phi(x,p_x), \kappa(y,p_y)] = i \delta^{(4)}(x-y)\delta^{(4)}(p_x-p_y), \nonumber \\
&\mbox{1st quantized: }& [\phi(x), \kappa(y)] = i \delta^{(4)}(x-y), \\
&\mbox{2nd quantized: }& [\Phi(x), \Pi(y)] = i\hbar c |\tau|\delta^{(4)}(x-y), \nonumber
\end{eqnarray}
which includes a time-like vector $\tau^{\mu}$ with dimensions of time to refer to the time-like displacement between spacetime foliations. An important insight was to realize that $\hat{\tau}^\mu$ also corresponds to the time-like component of a frame field. The simplest quantization procedure that we found is by setting $\Phi(x) = \phi(x)$ and $\Pi(x) = \hbar c |\tau| \kappa(x)$, where the extra factor of length $c|\tau|$ in comparison to $\vec{p} = \hbar \vec{k}$ is included to account for our use of four-dimensional delta functions, while canonical quantization typically leads to three-dimensional delta functions.

\begin{table}
\begin{tabular}{c||c|c}
Quantization Level & Field & Noncommutative Conjugate \\
\hline\hline
Classical phase space/0th quantized & $\phi(x,p)$, $\pi(x,p)$ & $\kappa(x,p)$, $\xi(x,p)$ \\
Classical spacetime/1st quantized & $\phi(x)$, $\pi(x)$ & $\kappa(x)$, $\xi(x)$ \\
2nd quantized & $\Phi(x)$ & $\Pi(x)$
\end{tabular}
\caption{The field content for classical, 1st quantized, and 2nd quantized scalar fields for the new covariant and symplectic approach are shown above. A field $A(a)$ with a noncommutative conjugate $B(b)$ satisfies $[A(a),B(b)] = i\delta(a-b)$. While KvN mechanics provides a wavefunction $\psi(x,p)$, the study of classical fields over phase space (0th quantized) such as $\phi(x,p)$ has been relatively unexplored. Classical fields such as $\phi(x)$ found in the Klein-Gordon equation are referred to as 1st quantized fields, since they are taken as a function of spacetime and have classical equations of motion with $\hbar$. 2nd quantized quantum fields such as $\Phi(x)$ are found as deformations of the 1st quantized fields.}
\label{table:012}
\end{table}

\subsection{A covariant Hamiltonian density closer to Dirac's}



The Dirac canonical momentum of a field $\phi(x)$ is given by $\pi_D(x) = \frac{\partial \mathcal{L}}{\partial \dot{\phi}}$, while the DDW canonical momentum is $\pi^\mu(x) = \frac{\partial \mathcal{L}}{\partial \partial_\mu \phi}$. The canonical momentum has the advantage of intuitively being the same rank as the field itself with symplectic structure, similar to spacetime and energy-momentum, while the DDW canonical momentum has the advantage of being Lorentz covariant. The canonical Hamiltonian theory requires a spacetime foliation with timelike separated hypersurfaces, while the DDW theory obscures the interpretation of time evolution for a more general notion of spacetime evolution. As it turns out, two different paths of exploration led to this symplectic Hamiltonian density by pushing the poly-symplectic geometry of DDW closer to Dirac's Hamiltonian formulation.


First, Koopman-von Neumann dynamics demonstrates that Poisson brackets and commutation relations provide different roles in classical theory, which becomes more apparent when generalizing to poly-symplectic fields. The Liouville operator in KvN mechanics is analogous with the Hamiltonian in quantum mechanics, but the poly-Liouville operator $\hat{\mathfrak{L}}_\mu$ found in Section \eqref{polyKvN} is now a covector, while the Hamiltonian is a scalar. This motivated the search for an analogue of the poly-Liouville operator as a type of energy-momentum density vector. The initial goal was to find a scheme in which Dirac's Hamiltonian density could be found in the time-like component when choosing $\hat{\tau}^\mu=(1,0,0,0)$.

Taking inspiration from Koopman-von Neumann mechanics in a De Donder-Weyl formulation, we define a vectorial poly-Hamiltonian density with the operator $\hat{\tau}^{\mu}$ which reduces to Dirac's Hamiltonian in the zeroth component when $\hat{\tau}^{\mu}=(1,0,0,0)$ is chosen,
\begin{equation}
\mathcal{H}^{\mu}=\hat{\tau}^{\rho} \nu_{\rho} \pi^{\mu}-\hat{\tau}^{\mu} \hat{\mathcal{L}}
\end{equation}
The timelike component of this operator contains the canonical momentum $\pi^{0}$. If $\hat{\tau}^{\mu}$ is a time-like constant, then $\int d^3x \mathcal{H}^{0}$ is identical to the canonical Hamiltonian found within a particular inertial reference frame. One approach could be to multiply the DDW Hamiltonian density by $\hat{\tau}^\mu$, but this wouldn't lead to the Dirac Hamiltonian density in the zeroth component in the same manner. However, the covector energy-momentum density functional above is no longer a Legendre transform of $\mathcal{L}$. Nevertheless, a scalar energy density functional can be defined by contracting $\mathcal{H}_{new} = \hat{\tau}_\mu\mathcal{H}^\mu$.

Second, consider a Legengre transformation with a ``velocity field'' $\nu$ with the new conjugate momentum field $\pi$ giving Dirac's canonical momentum when $\hat{\tau}^\mu = (1,0,0,0)$, 
\begin{equation}
\nu = \hat{\tau}^\mu \partial_\mu\phi = \hat{\tau}^\mu \nu_\mu, \qquad \pi = \frac{\partial \mathcal{L}}{\partial \nu} = \hat{\tau}_\mu \frac{\partial \mathcal{L}}{\partial \nu_\mu}.
\end{equation}
By taking inspiration from the covariant DDW Hamiltonian and striving for the simplicity of Dirac's Hamiltonian with symplectic geometry, a new type of Hamiltonian density $\mathcal{H}_{new}$ can be found that is covariant, symplectic, and canonical
\begin{equation}
\mathcal{H}_{new} = \nu\pi - \mathcal{L} = \hat{\tau}^\mu\nu_\mu \hat{\tau}_\nu \pi^\nu - \mathcal{L} = \hat{\tau}_\mu \mathcal{H}^\mu
\end{equation}
As mentioned, it turns out that $\mathcal{H}_{new} = \mathcal{H}^\mu \hat{\tau}_\mu$. The Hamiltonian density above is a Legendre transformation of the Lagrangian density since $\pi = \frac{\partial\mathcal{L}}{\partial \hat{\tau}^\mu \nu_\mu} = \hat{\tau}_\mu \frac{\partial \mathcal{L}}{\partial \nu_\mu}$.

To get time evolution in a specific frame and to connect back to Dirac's canonical quantization, consider a time-slice vector $\hat{\tau}^{\mu}=\frac{\tau^{\mu}}{|\tau|}$ that is $(1,0,0,0)$ in the rest frame. This same concept of a time-like unit vector has been considered by Wald and Iyer as well as Rovelli and Vidotto \cite{Iyer:1994ys,Iyer:1995kg,Wald:1999wa,RovelliVidotto}. The poly-Liouville operator contracted with $\hat{\tau}^{\mu}$ gives time evolution for arbitrary inertial frames with coordinates $x^{\prime \mu}=\left(c t^{\prime}, \vec{x}^{\prime}\right)$,
\begin{equation}
i\frac{1}{c} \frac{\partial}{\partial t^{\prime}} \phi(x)=i \hat{\tau}^{\mu} \partial_{\mu} \phi(x)=\hat{\tau}^{\mu} \hat{\mathfrak{L}}_{\mu} \phi(x) = i\hat{\tau}^\mu \pi_\mu.
\end{equation}
While this equation in a sense relates time evolution to the conjugate poly-momentum, it does not dynamically evolve the scalar field with the Klein-Gordon field equations. This relates to the fact that Schr\"odinger and KvN dynamics are first-order in time, while Klein-Gordon is second order and DDW contains two first-order equations. Before presenting the full equations of motion, we will demonstrate that $\phi$ and $\pi$ are a symplectic phase space of fields.

The symplectic two-form is typically given by $\eta = dx \wedge dp$, motivating a $2d$-dimensional phase space for $d$-dimensional spacetime. To demonstrate that this new Hamiltonian formulation contains symplectic structure, recall how Poisson brackets for 2D phase space can be described as a dyadic differential operator in terms of a symplectic metric $\eta = \left(\begin{array}{cc} 0 & 1 \\ -1 & 0 \end{array}\right)$,
\begin{equation}
\{A, B\} = A \left( \begin{array}{cc} \overleftarrow{\frac{\partial}{\partial x}} & \overleftarrow{\frac{\partial}{\partial p}}\end{array}\right) \left(\begin{array}{cc} 0 & 1 \\ -1 & 0 \end{array}\right) \left(\begin{array}{c} \frac{\partial}{\partial x} \\ \frac{\partial}{\partial p} \end{array}\right)B = \frac{\partial A}{\partial x} \frac{\partial B}{\partial p} - \frac{\partial A}{\partial p} \frac{\partial B}{\partial x}.
\end{equation}
Generalizing to fields, the partial derivatives with respect to $\phi$ and $\pi$ lead to field-theoretic Poisson brackets with symplectic structure,
\begin{equation}
\{ A, B \}_{new} = A \left( \begin{array}{cc} \overleftarrow{\frac{\partial}{\partial \phi}} & \overleftarrow{\frac{\partial}{\partial \pi}}\end{array}\right) \left(\begin{array}{cc} 0 & 1 \\ -1 & 0 \end{array}\right) \left(\begin{array}{c} \frac{\partial}{\partial \phi} \\ \frac{\partial}{\partial \pi} \end{array}\right)B = \frac{\partial A}{\partial \phi} \frac{\partial B}{\partial \pi} - \frac{\partial A}{\partial \pi} \frac{\partial B}{\partial \phi}.
\end{equation}
Mapping from the poly-symplectic geometry of De Donder-Weyl theory to symplectic structure is found by contracting $\hat{\tau}^\mu$ with $\{A , B\}_\mu$ shown in Eq.~\eqref{polyPoisson}. 
Separate from the commutation relations for the poly-KvN algebra, the Poisson brackets describe how these symplectic fields lead to canonical relations,
\begin{equation}
\{\phi, \pi\}_{new} = 1, \qquad \{\phi, \phi\}_{new} = \{\pi, \pi\}_{new} = 0.
\end{equation}
Upon completing this work, we realized that 
Witten, Zuckerman, and Crnkovic have considered a symplectic current that is covariant \cite{Witten1986,Zuckerman,Witten1987}. 
Crnkovic clarified how a symplectic current was the variation of the field times the poly-momentum, except the relationship to De Donder-Weyl theory was not realized \cite{Crnkovic1987,Crnkovic1988}. Their symplectic form $\omega$ includes an integral over a spacelike hypersurface, similar to Dirac's canonical quantization, but the Poincare invariance of $\omega$ was expressed. In this manner, our formulation is quite similar, as the choice of $\hat{\tau}_\mu$ relates to a choice of spacelike hypersurfaces. Our formulation seems to more easily provide a Hamiltonian density for general field theories. 

The new set of Hamilton's equations are found to be 
\begin{eqnarray}
\hat{\tau}^\mu \partial_\mu \phi &=& \{\phi,\mathcal{H}_{new}\}_{new} = \frac{\partial\mathcal{H}_{new}}{\partial \pi}, \\ 
\hat{\tau}^\mu \partial_\mu \pi &=& \{\pi,\mathcal{H}_{new}\}_{new} = -\frac{\partial\mathcal{H}_{new}}{\partial\phi}. 
\end{eqnarray}
Note how these form of Hamilton's equations apply to arbitrary fields and their associated conjugate momentum fields, not just scalar fields. While these equations of motion are in terms of a new kind of Poisson brackets, the fields $\phi$ and $\pi$ can be found in a commutator algebra that generalizes the KvN algebra to fields.

Since the KvN algebra contains Fourier conjugate variables of phase space such as $\vec{k}$ and $\vec{q}$ as shown in Eq.~\eqref{KvNalg}, there should also be generalized Fourier conjugate fields $\kappa(x)$ and $\xi(x)$ that are conjugate to $\phi(x)$ and $\pi(x)$. Additionally, a field $\phi(x)$ is in a sense a generalization of a quantum wavefunction $\psi(x)$, while the classical KvN wavefunction $\psi(x,p)$ depends both on spacetime and energy-momentum. For this reason, generalizing KvN mechanics to fields implies that the fully classical or 0th-quantized fields may be a function over all of phase space. From this perspective, classical field theory typically studies 1st-quantized fields, which describes how the massive Klein-Gordon equation can be studied as a classical field theory, yet still contain $\hbar$, as the coordinates are quantized, but not the fields themselves. 

The 0th-quantized KvN algebra associated with the fields $\phi(x,p)$, $\pi(x,p)$, $\kappa(x,p)$, and $\xi(x,p)$ is given by the following commutation relations,
\begin{eqnarray}
[\phi(x,p_x), \kappa(y,p_y)] &=& i\delta^{(4)}(x-y) \delta^{(4)}(p_x-p_y), \\ 
{[}\xi(x,p_x), \pi(y,p_y)] &=& i\delta^{(4)}(x-y) \delta^{(4)}(p_x-p_y).
\end{eqnarray}
The ``position field basis'' for 0th quantized fields are
\begin{eqnarray}
\kappa(x,p_x) &=& -i\delta^{(4)}(x-y) \delta^{(4)}(p_x - p_y) \frac{\partial}{\partial \phi(y,p_y)}, \\ 
\xi(x,p_x) &=& i\delta^{(4)}(x-y) \delta^{(4)}(p_x - p_y) \frac{\partial}{\partial \pi(y,p_y)}.
\end{eqnarray}

The 1st-quantized algebras associated with these fields $\phi(x)$, $\pi(x)$, $\kappa(x)$, and $\xi(x)$ gives
\begin{eqnarray}
[\phi(x), \kappa(y)] &=& i\delta^{(4)}(x-y) , \\ 
{[}\xi(x), \pi(y)] &=& i\delta^{(4)}(x-y) .
\end{eqnarray}
The ``position field basis'' for 0th and 1st quantized fields are
\begin{eqnarray}
\kappa(x) &=& -i\delta^{(4)}(x-y) \frac{\partial}{\partial \phi(y)}, \\ 
\xi(x) &=& i\delta^{(4)}(x-y)  \frac{\partial}{\partial \pi(y)}.
\end{eqnarray}
Alternatively, a ``wavenumber field basis'' could be chosen that finds $\phi(x) = i \delta^{(4)}\frac{\partial}{\partial \kappa(y)}$. 

Next, we demonstrate that the interacting Klein-Gordon, Maxwell, and linearized gravity equations of motion can all be derived from this new Hamiltonian density. 

\subsubsection{Covariant Hamiltonian density for Klein-Gordon scalars}

To derive the new Hamiltonian density, it is convenient to introduce a frame field $e_a^\mu = \left(\hat{\tau}^\mu, \hat{x}^\mu, \hat{y}^\mu, \hat{z}^\mu\right)$. For non-gravitational theories, $e_a^\mu$ as a constant simply encodes Lorentz transformations between the global Minkowski manifold in some frame and local coordinates that can be in a different frame. The local and global frames are equivalent when $e_a^\mu$ is given by the identity matrix, which leads to $\hat{\tau}^\mu = (1,0,0,0)$. 

The Klein-Gordon action can be written as
\begin{equation}
S = \int d^4x \left[ \frac{1}{2} \hat{\tau}^\mu \partial_\mu \phi \hat{\tau}_\nu \partial^\nu \phi + \frac{1}{2} e_i^\mu \partial_\mu \phi e^i_\nu \partial^\nu \phi - V(\phi)\right], 
\end{equation}
where $i=1,2,3$. The velocity field and conjugate momentum field are 
\begin{equation}
\nu = \hat{\tau}^\mu \partial_\mu \phi, \qquad \pi = \frac{\partial\mathcal{L}}{\partial \nu} = \hat{\tau}_\mu\partial^\mu \phi.
\end{equation}

The new Hamiltonian density for an interacting Klein-Gordon field is
\begin{equation}
\mathcal{H}_{new} = \frac{1}{2}\left(\pi^2 - e_i^\mu\partial_\mu \phi e_\nu^i \partial^\nu \phi \right) + V(\phi).
\end{equation}
Integrating by parts for the second term and dropping the boundary term allows for an easier evaluation of Hamilton's equations,
\begin{equation}
\mathcal{H}_{new} = \frac{1}{2}\left(\pi^2 + \phi  e_i^\mu e_\nu^i \partial_\mu\partial^\nu \phi \right) + V(\phi) + \mbox{bdry},
\end{equation}
where we assume that $e_i^\mu$ only can specify local inertial frames that correspond to global Lorentz transformations on Minkowski spacetime. 
Applying the new set of Hamilton's equations gives
\begin{eqnarray}
\hat{\tau}^\mu\partial_\mu \phi &=& \{\phi,\mathcal{H}_{new}\}_{new} = \pi, \\ 
\hat{\tau}^\mu\partial_\mu \pi &=& \{\pi,\mathcal{H}_{new}\}_{new} = - e_i^\mu e^i_\nu \partial_\mu\partial^\nu \phi - \frac{\partial V}{\partial\phi}.
\end{eqnarray}
Plugging the first equation into the second leads to the Klein-Gordon equation of motion. In this manner, the Hamiltonian dynamics are identical to Dirac's Hamiltonian, except the frame fields are introduced to allow for arbitrary local frames. The new Hamiltonian density found is truly covariant. Understanding the role of $e_a^\mu$ as a frame field also allows for a more dynamical realization of foliations. Despite $\hat{\tau}^\mu$ being the zeroth component of $e_a^\mu$, the local structure is independent of the global manifold. In this manner, Dirac's Hamiltonian is not covariant with respect to the global manifold since $\mu=0$ is isolated. The frame field allows for $a=0$ to be chosen locally in a manner that the Hamiltonian density is still Lorentz covariant with respect to the global manifold.

\subsubsection{Covariant Hamiltonian density for Maxwell theory}

The Maxwell action is
\begin{equation}
S = \int d^4x \mathcal{L} =  \int d^4x \left( -\frac{1}{4}F_{\mu\nu} F^{\mu\nu} - A^\mu J_\mu \right). 
\end{equation}
The velocity field of interest is 
\begin{equation}
\nu_\mu = \hat{\tau}^\nu \partial_\nu A_\mu.
\end{equation}
The following relation allows for the Lagrangian to be expressed in terms of the velocity field,
\begin{equation}
\delta^\mu_\nu = e^\mu_a e^a_\nu = \hat{\tau}^\mu \hat{\tau}_\nu + e^\mu_i e^i_\nu.\label{delta}
\end{equation}
The Lagrangian density can be expanded to give
\begin{eqnarray}
\mathcal{L} &=& -\frac{1}{4}F_{\mu\nu}F_{\rho\sigma} \left( \hat{\tau}^\mu \hat{\tau}^\nu \hat{\tau}^\rho \hat{\tau}^\sigma + \hat{\tau}^\mu e^\nu_j \hat{\tau}^\rho e^{\sigma j} + e^\mu_i \hat{\tau}^\nu e^{\rho i} \hat{\tau}^\sigma + e^\mu_i e^\nu_j e^{\rho i} e^{\sigma j}\right) \\
&=& -\frac{1}{2} e^\nu_j \nu^\nu e^{\sigma j} \nu_\sigma + e^{\sigma j} \nu_\sigma  e^\nu e^\nu_j \hat{\tau}^\mu \partial_\nu A_\mu -\frac{1}{2} e^\nu_j \hat{\tau}^\mu \partial_\nu A_\mu e^{\sigma j} \hat{\tau}^\rho \partial_\sigma A_\mu -\frac{1}{4} F_{\mu\nu}F^{\rho\sigma}e^\mu_i e^\nu_j e_\rho^i e_\sigma^j. \nonumber
\end{eqnarray}

The conjugate momentum field is found to be
\begin{equation}
\pi^\sigma = \frac{\partial\mathcal{L}}{\partial \nu_\sigma} = -e^\nu_j e^{\sigma j} \left(\nu_\nu - \hat{\tau}^\mu \partial_\nu A_\mu\right) = - e^\nu_j e^{\sigma j} \hat{\tau}^\mu F_{\mu\nu}. 
\end{equation}
Another way to express the conjugate momentum field is given by
\begin{equation}
\hat{\tau}^\nu F_{\mu\nu} = \left( \hat{\tau}^\rho \hat{\tau}_\mu + e^\rho_i e_\mu^i \right) F_{\rho\nu}\hat{\tau}^\nu = \hat{\tau}_\mu \hat{\tau}^\rho \hat{\tau}^\nu F_{\rho\nu} + \pi_\mu = \pi_\mu, 
\end{equation}
where the antisymmetry of $F_{\rho\nu}$ led to the vanishing of the first term above. A background-independent derivation using differential forms would automatically impose this antisymmetry. Both forms of the conjugate momentum allow for the Lagrangian density to be rewritten as
\begin{equation}
\mathcal{L} = -\frac{1}{2} \pi^\mu \pi_\mu - \frac{1}{4} F_{\mu\nu}F^{\rho\sigma}e^\mu_i e^\nu_j e_\rho^i e_\sigma^j - A^\mu J_\mu.
\end{equation}
The covariant Hamiltonian density is found as
\begin{equation}
\mathcal{H}_{new} = \nu^\mu \pi_\mu - \mathcal{L} = \pi_\mu\left( -\frac{1}{2} \pi^\mu + \hat{\tau}_\nu \partial^\mu A^\nu\right) + \frac{1}{4} F_{\mu\nu}F^{\rho\sigma}e^\mu_i e^\nu_j e_\rho^i e_\sigma^j + A^\mu J_\mu.
\end{equation}
Covariant Hamilton's equations (after integrating by parts) lead to
\begin{eqnarray}
\hat{\tau}^\mu \partial_\mu A_\nu &=& \frac{\partial \mathcal{H}_{new}}{\partial \pi^\nu} = -\pi_\nu + \hat{\tau}^\mu \partial_\nu A_\mu, \\
\hat{\tau}^\mu \partial_\mu \pi_\nu &=& - \frac{\partial \mathcal{H}_{new}}{\partial A^\nu} = \partial^\mu \pi_\mu \hat{\tau}_\nu + \partial^\mu F_{\rho\sigma} e_\mu^i e_\nu^j e^\rho_i e^\sigma_j - J_\nu. 
\end{eqnarray}
Plugging the first equation into the second leads to 
\begin{eqnarray}
J_\nu &=&  -\hat{\tau}^\mu \partial_\mu \pi_\nu + \partial^\mu\pi_\mu \hat{\tau}_\nu + \partial^\mu F_{\rho\sigma}e_\mu^i e_\nu^j e^\rho_i e^\sigma_j \nonumber \\ 
&=& e_\nu^i e^\sigma_i \hat{\tau}_\mu \hat{\tau}^\rho \partial^\mu F_{\rho\sigma} + \hat{\tau}^\rho \hat{\tau}_\nu \partial^\mu F_{\mu\rho} + \partial^\mu F_{\rho\sigma}e_\mu^i e_\nu^j e^\rho_i e^\sigma_j \nonumber \\ 
&=& \partial^\mu F_{\rho\sigma} \left(\hat{\tau}_\mu \hat{\tau}^\rho e_\nu^i e^\sigma_i + \hat{\tau}^\sigma \hat{\tau}_\nu e^\sigma_i e_\mu^i + e_\mu^i e_\nu^j e^\rho_i e^\sigma_j\right) \nonumber \\
&=& \partial^\mu F_{\mu\nu}.
\end{eqnarray}
where using Eq.~\eqref{delta} and $\hat{\tau}^\rho\hat{\tau}^\sigma F_{\rho\sigma} = 0$ were used to simplify the result above.

\subsubsection{Covariant Hamiltonian density for linearized gravity}

Since general relativity would require a dynamical frame field $e_\mu^i(x)$ that is position dependent, it is worthwhile to first establish the new Hamiltonian density for linearized gravity. The linearized gravity action with matter is 
\begin{equation}
S = \int d^4x \left(\frac{1}{2} \partial^\rho\bar{h}^{\mu\nu} \partial_\rho \bar{h}^{\mu\nu} -\frac{1}{4}\partial_\mu\bar{h}\partial^\mu \bar{h} \right)
\end{equation}
Treating $\bar{h}_{\mu\nu}$ as the dynamical field leads to the velocity field $\bar{\nu}_{\mu\nu}$,
\begin{equation}
\bar{\nu}_{\mu\nu} = \hat{\tau}^\rho\partial_\rho \bar{h}_{\mu\nu}.
\end{equation}
The Lagrangian can be rewritten as
\begin{equation}
\mathcal{L} = \frac{1}{2}\left(\bar{\nu}_{\mu\nu}\bar{\nu}^{\mu\nu} -\frac{1}{2}\bar{\nu}\bar{\nu} + e_\rho^i e^\sigma_i\left(\partial^\rho \bar{h}^{\mu\nu} \partial_\sigma\bar{h}_{\mu\nu} -\frac{1}{2} \partial^\rho\bar{h} \partial_\sigma\bar{h}\right) \right) -\frac{8\pi G}{c^4}\kappa \hat{T}^{\mu\nu}\hat{h}_{\mu\nu},
\end{equation}
where $\bar{\nu} \equiv \bar{\nu}_{\mu\nu}\eta^{\mu\nu}$. 
The conjugate momentum field is found to be
\begin{equation}
\bar{\pi}^{\mu\nu} = \frac{\partial \mathcal{L}}{\partial \bar{\nu}_{\mu\nu}} = \bar{\nu}^{\mu\nu} - \frac{1}{2}\eta^{\mu\nu} \bar{\nu} = \nu^{\mu\nu} = \hat{\tau}_\rho \partial^\rho h^{\mu\nu},
\end{equation}
where a reciprocal relationship is found such that $\nu_{\mu\nu} = \bar{\pi}_{\mu\nu}$, etc. 

The  covariant Hamiltonian density in terms of the gravitational field $\bar{h}_{\mu\nu}$ and conjugate momentum field $\bar{\pi}^{\mu\nu}$ is
\begin{equation}
\mathcal{H}_{new} = \frac{1}{2} \bar{\pi}_{\mu\nu}\bar{\pi}^{\mu\nu} -\frac{1}{4} \bar{\pi}\bar{\pi} - \frac{1}{2}e_\rho^i e^\sigma_i\left(\partial^\rho \bar{h}^{\mu\nu} \partial_\sigma\bar{h}_{\mu\nu} -\frac{1}{2} \partial^\rho\bar{h} \partial_\sigma\bar{h}\right) + \frac{8\pi G}{c^4}\kappa \bar{T}^{\mu\nu}\bar{h}_{\mu\nu}. 
\end{equation}
The covariant Hamilton's equations give
\begin{eqnarray}
\hat{\tau}^\rho\partial_\rho \bar{h}_{\mu\nu} &=& \frac{\partial \mathcal{H}_{new}}{\partial \bar{\pi}^{\mu\nu}} = \bar{\pi}_{\mu\nu} - \frac{1}{2}\eta_{\mu\nu}\bar{\pi} = \pi_{\mu\nu}, \\ 
\hat{\tau}^\rho \partial_\rho \bar{\pi}_{\mu\nu} &=& - \frac{\partial\mathcal{H}_{new}}{\partial \bar{h}^{\mu\nu}} = -\left(e_\rho^i e^\sigma_i \partial^\rho \partial_\sigma \left(\bar{h}_{\mu\nu}- \frac{1}{2}\eta_{\mu\nu} \bar{h} \right) + \frac{8 \pi G}{c^4}\kappa \bar{T}^{\mu\nu}\right).
\end{eqnarray}
Plugging the first equation in the trace reversal of the second equation and rearranging leads to
\begin{equation}
\left( \hat{\tau}_\rho  \hat{\tau}^\sigma + e_\rho^i e^\sigma_i\right) \partial^\rho\partial_\sigma \bar{h}_{\mu\nu} = \partial^\rho\partial_\rho \bar{h}_{\mu\nu} = - \frac{8\pi G}{c^4}\kappa T^{\mu\nu}.
\end{equation}
This reproduces the known equations of motion for linearized gravity. 



\subsection{Quantization of the new Hamiltonian theory}

Using the covariant and canonical Hamiltonian density $\mathcal{H}_{new}$, four sets of classical fields $\phi(x)$, $\kappa(x)$, $\pi(x)$, and $\xi(x)$ can be used to construct two quantum fields $\Phi(x)$ and $\Pi(x)$,
\begin{eqnarray}
\Phi(x) &=& a \phi(x) + b \xi(x), \\ 
\Pi(x) &=& c\pi(x) + d \kappa(x).
\end{eqnarray}
The commutation relations of the second-quantized fields are therefore
\begin{equation}
[\Phi(x), \Pi(y)] = \left( ad + bc \right) i \delta^{(4)}(x-y).
\end{equation}
Setting $a=1$, $d =\hbar c|\tau|$ and $b=c=0$ is the simplest quantization procedure, although other possibilities exist, as mentioned in Eq.~\eqref{solns}. 

The following commutation relations are found with $\Pi(x) = \hat{\tau}^\mu \Pi_\mu(x)$ as the symplectic, covariant, and canonical momentum, 
\begin{eqnarray}
[\Phi(x), \Pi(y)] &=& i\hbar c|\tau|\delta^{(4)}(x-y), \nonumber \\
{[}\Phi(x), \Phi(y)] &=& [\Pi(x),\Pi(y)] = 0. 
\end{eqnarray}
Similar to how $p_\mu = \hbar k_\mu$, it is anticipated that $\Pi(x) = \hbar c|\tau| \kappa(x)$.



If one were to model the universe as a quantum computer simulation, then this quantum computer must simulate the universe at a fast enough rate such that all possible observables lead to self-consistent results. Since the Planck time is thought to be the smallest possible duration of time that could be measured, it is sensible to assume that such a quantum computer simulation would use spacelike foliations separated by times no larger than the Planck time. Thiemann states that spin quantum numbers in LQG are created and destroyed in a Planck moment and Zizzi has discussed Planck-time foliations \cite{Thiemann,Zizzi:2018smh}. For our construction, this implies that $|\tau| = t_P = \sqrt{\frac{\hbar G}{c^5}}$. With this assumption, the new canonical, covariant, and symplectic commutation relations for fields in natural units with $c=\hbar = G = 1$, 
\begin{equation}
[\Phi(x), \Pi(y)] = i\delta^{(4)}(x-y). \label{4DCommRel}
\end{equation}
This canonical commutation relation for second quantized fields appears to be the simplest possible that are manifestly covariant and symplectic. Building off of earlier work \cite{Diaz:2018uny,Diaz:2019xie,Diaz:2020dfe}, Eq.~\eqref{4DCommRel} was recently presented in Ref.~\cite{Diaz:2021snw} by different motivations. 


\section{Conclusions}

In this work, we have demonstrated that a Hamiltonian density for relativistic field theory can be found that is canonical, covariant, and symplectic. This was found by taking the De Donder-Weyl poly-momentum field and contracting with the time-like component of a frame field to obtain a symplectic momentum field. This momentum field can locally interpolate between different ADM foliations of spacetime, which makes it closer to the Dirac's canonical momentum. The generalized Hamilton's equations for this new Hamiltonian density were found and demonstrated to give the correct equations of motion for Klein-Gordon, Maxwell, and linearized gravity theories. We also generalized the phase space of fields to a Fourier-phase space of fields as a generalization of Koopman-von Neumann mechanics, giving a classical commutator algebra of fields. KvN quantization of these new fields was provided, which bypasses the use of Poisson brackest to map from classical commutators to quantum commutators. 

In Appendix \eqref{KvN}, a relativistic formulation of Koopman-von Neumann classical mechanics was constructed, which found a problem of time for the proper Liouville operator for the point particle action. The poly-symplectic geometry of De Donder-Weyl theory was introduced in Appendix \eqref{DDW} to study the generalization of relativistic Koopman-von Neumann mechanics to field theory, which allows for the classical equations of motion to be derived quickly from the De Donder-Weyl Hamiltonian density. To our knowledge, classical, 1st, and 2nd quantized poly-Koopman-von Neumann algebras for fields were presented for the first time, which was shown to be compatible with De Donder-Weyl dynamics. 



In future work, it would be worthwhile to derive the general Yang-Mills field equations and Einstein's field equations from their respective covariant, canonical, and symplectic Hamiltonian densities. The introduction of the frame field for specifying local frames requires additional care in general relativity. Since the frame field is the gauge field of the translations, formulations of gauge gravity such as subsectors of metric-affine gauge gravity should be explored. Another avenue of exploration could use the relativistic Koopman-von Neumann algebra introduced here to extend the prequantum operator algebra defined in \cite{opmech2022} and its possible category-theoretic realizations using pregeometric constructions discussed in \cite{pregeo1,pregeo2}. 

Additionally, it would be worthwhile to explore singular Lagrangian densities, such as spinor fields. Other work has attempted to use a quadratic action with a constraint to describe spinors \cite{Torres-Gomez:2012tka}. Revisiting the Dirac-Bergmann algorithm for singular Lagrangians would also be appropriate \cite{HansonRegge}. Also, Kanatchikov has introduced a generalization of the Dirac bracket formula to DDW theory for degenerate Lagrangian densities, such as the Dirac Lagrangian density discussed above \cite{Kanatchikov:2008he}. Finally, our work may be inspirational for a covariant and canonical formulation of loop quantum gravity.





\begin{acknowledgements}
Thanks to Peter Morgan for making us aware of Koopman-von Neumann mechanics \cite{Morgan:2019azd,Morgan:2021jeh}. Thanks to Marcelo Amaral and Ray Aschheim for discussions related to canonical quantization and De Donder-Weyl theory. Thanks to Carlos Castro-Perelman for various discussions related to phase space. 
Thanks to Igor Mol and Richard Clawson for helpful comments and feedback.
\end{acknowledgements}

\appendix

\section{Koopman-von Neumann mechanics}
\label{KvN}

Rather than reviewing KvN mechanics, we introduce a relativistic KvN framework using $\hat{x}^\mu$, $\hat{p}_\nu$, $\hat{k}_\rho$, and $\hat{q}^\sigma$, leading to an appropriate Liouville operator that determines the proper time evolution. We find that the problem of time occurs in this Koopman von-Neumann classical mechanics for the relativistic point particle action. This suggests that the problem of time is not a problem with quantum mechanics, but rather stems from the choice of the appropriate Hamiltonian formulation. Nevertheless, the KvN quantization of special relativity can also be found by imposing $\hat{p}_\mu = \hbar \hat{k}_\mu$, which is not generally true in classical mechanics. 

\subsection{Classical relativistic Koopman-von Neumann mechanics}
\label{relKvN}

We start by introducing classical special relativity within Koopman-von Neumann (KvN) mechanics. For an introduction to KvN mechanics, see Ref.~\cite{Piasecki:2021keq}. The two-state-vector formalism interpretation of quantum mechanics has been related to the entangled histories interpretation \cite{Watanabe,Aharonov1964,Aharonov1991,Aharonov2009,CotlerWilczek,Nowakowski:2016lvd,Nowakowski2017,Nowakowski:2018nvh}. Entangled histories allows for a Hilbert space for time, which has also been discussed elsewhere \cite{Giovannetti:2015qha,Maccone,Giovannetti:2022pab,Vaccaro:2018lvi,Diaz:2018uny,Diaz:2019xie,Diaz:2020dfe}. To the best of our knowledge, a Hilbert space for time has never been discussed within KvN mechanics, which helps provide a relativistic formulation of Koopman-von Neumann mechanics. 

The standard Dirac bra-ket notation can be used for classical mechanics by admitting a wavefunction $\psi(A)=\langle A | \psi\rangle$, and $\psi^{*}(A)=\langle\psi | A\rangle$. A basis for phase space in classical mechanics is provided with $|A\rangle=|x, p\rangle=|x\rangle \otimes|p\rangle$, since position and momentum are independent. Unlike QM, the spacetime position operator $\hat{x}^{\mu}$ commutes with the energy-momentum operator $\hat{p}_{\nu}$ when acting on $\psi(x, p)$,
\begin{equation}
\left[\hat{x}^{\mu}, \hat{p}_{\nu}\right] \psi(x, p)=0 .
\end{equation}
While classical non-relativistic phase space allows for $\psi(\vec{x}, \vec{p}, t)$ as seven dimensions, the relativistic generalization initially motivates 9 dimensions (8 independent) via $\psi\left(x^{\mu}, p_{\nu}, \tau\right)$. 

To provide proper time evolution of a Hilbert space in a classical setting, a proper Hamiltonian $H$ is introduced from a proper Lagrangian $L$, such that a canonical and covariant action is
\begin{equation}
S = \int d\tau L = \int d\tau (v^\mu p_\mu - H),
\end{equation}
where $v^\mu = \frac{\partial x^\mu}{\partial \tau}$ is the relativistic 4-velocity and $p_\mu = \frac{\partial L}{\partial v^\mu}$ is the canonical energy-momentum. Proper Hamilton's equations are found by varying about the classical path, giving
\begin{eqnarray}
\frac{\partial x^\mu}{\partial \tau} = \frac{\partial H}{\partial p_\mu} = \{x^\mu ,H\}_{PB}, \\ 
\frac{\partial p_\mu}{\partial \tau} = - \frac{\partial H}{\partial x^\mu} = \{ p_\mu ,H\}_{PB}
\end{eqnarray}
where the relativistic Poisson brackets are given by 
\begin{equation}
\{A, B\}_{P B} = \frac{\partial A}{\partial x^{\mu}} \frac{\partial B}{\partial p_{\mu}}-\frac{\partial A}{\partial p_{\mu}}\frac{\partial B}{\partial x^{\mu}}.
\end{equation}

The proper Hamiltonian $H$ allows for a proper relativistic Liouville equation in terms of the probability density $\rho(x, p, \tau)$,
\begin{equation}
\frac{d \rho}{d \tau}= \frac{\partial\rho}{\partial\tau} + \frac{\partial x^\mu}{\partial\tau}\frac{\partial\rho}{\partial x^\mu} + \frac{\partial p_\mu}{\partial \tau} \frac{\partial \rho}{\partial p_\mu} =\frac{\partial \rho}{\partial \tau}+\frac{\partial H}{\partial p_{\mu}} \frac{\partial \rho}{\partial x^{\mu}}-\frac{\partial H}{\partial x^{\mu}} \frac{\partial \rho}{\partial p_{\mu}} = \frac{\partial \rho}{\partial \tau} + \{\rho, H\}_{PB} = \frac{\partial \rho}{\partial \tau}+i \hat{\mathfrak{L}} \rho=0 .
\end{equation}
where $\hat{\mathfrak{L}}$ is the proper Liouville operator, which is Hermitian in a manner similar to the Hamiltonian operator $\hat{H}$ in quantum mechanics. The proper Liouville operator is found from the proper Liouville equation after applying Hamilton's equations. The proper Liouville operator can be expressed with relativistic Poisson brackets, giving
\begin{equation}
i \hat{\mathfrak{L}} A = \{A, H\}_{P B} = \frac{\partial A}{\partial x^{\mu}} \frac{\partial H}{\partial p_{\mu}}-\frac{\partial A}{\partial p_{\mu}}\frac{\partial H}{\partial x^{\mu}}. 
\end{equation}
The proper Liouville operator acting on $x^\mu$ and $p_\mu$ gives proper Hamilton's equations, while acting on $\rho$ gives the proper Liouville equation. 

Koopman-von Neumann mechanics expresses probability densities $\rho(x,p) = \psi^*(x,p) \psi(x,p)$ in terms of the Koopman-von Neumann wavefunction $\psi(x,p)$. While relativistic field theory typically defines a probability density of a quantum field by taking a time derivative after generalizing the current to a 4-current, the Klein-Gordon field equation has one additional time derivative not found in the Schr\"odinger equation. To obtain a relativistic analogue of the KvN equations, we assume $\rho(x, p, \tau)=\psi^{*}(x, p, \tau) \psi(x, p, \tau)=|\langle x, p \mid \psi(\tau)\rangle|^{2}$. 
The proper Liouville operator is used to lead to the proper Koopman-von Neumann equation,
\begin{equation}
i \frac{\partial \psi}{\partial \tau}=\hat{\mathfrak{L}} \psi. \label{properKvN}
\end{equation}
This equation is a classical analogue of the Schr\"odinger equation (generalized to a relativistic setting with proper time evolution). 

In summary, classical KvN mechanics has four axioms
\begin{enumerate}
  \item The state of the system is representated by $|\psi\rangle$ in a complex Hilbert space.
  \item An observable is a Hermitian operator $\hat{A}$ for an eigenvalue $A$ with eigenstate $|A\rangle$ satisfying $\hat{A}|\psi\rangle=A|\psi\rangle$.
  \item The probability of $A$ is given by $P(A)=|\langle A \mid \psi(t)\rangle|^{2}$ as the Born rule, which leads to instantaneous collapse of the wavefunction. 
  \item The tensor product of subsystems leads to a description of the composite system.
\end{enumerate}
The third axiom should be questioned for a relativistic formulation of classical mechanics, as retarded functions for information transfer of measurements should be incorporated. For now, our interpretation is that the wavefunction for observers infinitesimally close to the measurement should observe the collapse of the wavefunction within an infinitesimal amount of time.

The Poisson brackets of $x^\mu$ and $p_\nu$ describe the symplectic structure of phase space, since
\begin{eqnarray}
\{x^\mu, p_\nu \}_{PB} &=& \delta^\mu_\nu, \\ 
\{x^\mu, x^\nu\}_{PB} &=& \{p_\mu,p_\nu\}_{PB} = 0. 
\end{eqnarray}
Groenewold showed that mapping classical Poisson brackets to quantum commutators as suggested by Dirac is not general \cite{Groenewold}. The quantization procedure we pursue throughout relates more to $\hat{p}_\mu = \hbar \hat{k}_\mu$, which can be understood more clearly from the Koopman-von Neumann algebra. 
The KvN algebra is given by
\begin{eqnarray}
\left[\hat{x}^{\mu}, \hat{k}_{\nu}\right] &=& i \delta_{\nu}^{\mu}, \label{KvNalg}\\
\left[\hat{q}^{\mu}, \hat{p}_{\nu}\right] &=& i \delta_{\nu}^{\mu}, \\
\left[\hat{x}^{\mu}, \hat{p}_{\nu}\right] &=& \left[\hat{x}^{\mu}, \hat{q}^{\nu}\right]=\left[\hat{p}_{\nu}, \hat{k}_{\nu}\right]=\left[\hat{q}^{\mu}, \hat{k}_{\nu}\right]=0, \\
\left[\hat{x}^{\mu}, \hat{x}^{\nu}\right] &=& \left[\hat{p}_{\mu}, \hat{p}_{\nu}\right]=\left[\hat{k}_{\mu}, \hat{k}_{\nu}\right]=\left[\hat{q}^{\mu}, \hat{q}^{\nu}\right]=0 .
\end{eqnarray}
The KvN algebra is a commutative tensor product of two Heisenberg algebras (without any notion of $\hbar$). 
The KvN algebra establishes that we have both ${A,B}$ and $[A,B]$ in the classical setting. Rather than applying ${A,B} \rightarrow \frac{1}{i\hbar}[A,B]$, rigorous quantization can be found by deforming the KvN algebra \cite{Bondar,opmech2022}. We updated the notation to reflect that $\hat{k}_\mu$ is a frequency-wavenumber operator conjugate to $\hat{x}^\mu$, wihle $\hat{q}^\mu$ is a frequency-wavenumber operator conjugate to energy-momentum $\hat{p}_\mu$.  As such, the KvN algebra relates deeply to the structure of Fourier-phase space.

Natural representations for the wavenumber operators $\hat{k}_{\nu}$ and $\hat{q}^{\mu}$ in the phase space basis $|A\rangle = |x,p\rangle$ are
\begin{equation}
\hat{k}_{\nu}=-i \partial_{\nu} \equiv-i \frac{\partial}{\partial x^{\nu}}, \quad \hat{q}^{\mu}=i \tilde{\partial}^{\mu} \equiv i \frac{\partial}{\partial p_{\mu}} .
\end{equation}
The operators $\hat{x}^{\mu}$ and $\hat{p}_{\mu}$ lead to the following eigenvalues when acting on a phase space eigenstate,
\begin{equation}
\hat{x}^{\mu}|x, p\rangle=x^{\mu}|x, p\rangle, \quad \hat{p}_{\mu}|x, p\rangle=p_{\mu}|x, p\rangle .
\end{equation}
In the classical Koopman-von Neumann formulation, there is also a Fourier space basis $|A\rangle=|k, q\rangle$. In this basis, the operators can be represented with a bar, yet they must satisfy the same Koopman-von Neumann algebra in Eq. (8). The position and momentum operators become differential operators in the Fourier-conjugate basis, 
\begin{equation}
\hat{\bar{x}}^{\mu}=i \bar{\partial}^{\mu}=i \frac{\partial}{\partial k_{\mu}}, \quad \hat{\bar{p}}_{\mu}=-i \overline{\tilde{\partial}}_{\mu}=-i \frac{\partial}{\partial q^{\mu}} .
\end{equation}
The wavenumber operators of phase space act on wavenumber eigenstates to return wavenumber eigenvalues,
\begin{equation}
\hat{\bar{k}}_{\mu}|k, q\rangle=k_{\mu}|k, q\rangle, \quad \hat{\bar{q}}^{\mu}|k, q\rangle=q^{\mu}|k, q\rangle .
\end{equation}

While this formulation leads to a manifestly relativistic KvN equation in terms of proper time, it leads to a problem of time for the Lagrangian density of a free point particle, as the proper Liouville operator leads to zero. Consider the action for a point particle in curved spacetime given by the proper length,
\begin{equation}
S = -mc\int d\tau \sqrt{- g_{\mu\nu} \frac{dx^\mu}{d\tau}\frac{dx^\nu}{d\tau}}.
\end{equation}
The proper Lagrangian $L$ is given by $S = \int d\tau L$, which leads to a proper covariant momentum,
\begin{equation}
p_\mu = \frac{\partial L}{\partial v^\mu} = mc g_{\mu\nu} v^\nu \frac{1}{\sqrt{-g_{\alpha\beta}v^\alpha v^\beta}} = m v_\mu,
\end{equation}
where $v^\mu = \frac{d x^\mu}{d\tau}$ and $v^\mu v_\mu = -c^2$ was used to find the result above. This allows for the solution of the proper Hamiltonian $H$ to be found,
\begin{equation}
H = v^\mu p_\mu - L = mv^\mu v_\mu + mc\sqrt{-g_{\mu\nu}v^\mu v^\nu} = -mc^2 + mc^2 = 0.
\end{equation}
In the quantum theory, the proper Hamiltonian density operator would also be zero. 

A similar problem occurs with the Wheeler-DeWitt equation, except the zero Hamiltonian is interpretted as a constraint \cite{DeWitt:1967yk,Isham:1992ms}. With relativistic Koopman-von Neumann mechanics, the proper Hamiltonian would be zero, which leads to a zero Liouville operator for time evolution of phase space. This suggests that the problem of time is independent of quantum theory, but rather is a problem with some Hamiltonian systems for relativistic theories. To avoid this problem and keep the manifest Lorentz invariance, De Donder-Weyl theory is pursued next, rather than a proper Hamiltonian formulation. 





\subsection{Quantization of relativistic Koopman-von Neumann mechanics}
While classical theory treats $p_{\mu}$ and $k_{\mu}$ as independent, quantum mechanics treats the operators as linearly dependent via $\hat{P}_{\mu}=\hbar \hat{K}_{\mu}$ and $\hat{Q}^{\mu}=\hbar \hat{X}^{\mu}$. This relates to the Heisenberg uncertainty relations of spacetime and energy-momentum in the relativistic setting. From the operator algebra perspective, quantization of classical theory can be seen in a straightforward manner with Koopman-von Neumann mechanics, as both utilize a Hilbert space. A quantum deformation of classical phase space can be implemented as a change of variables from $\hat{x}^{\mu} \rightarrow \hat{X}^{\mu}$ and $\hat{p}_{\nu} \rightarrow \hat{P}_{\nu}$
\begin{equation}
\left[\hat{x}^{\mu}, \hat{p}_{\nu}\right]=0 \rightarrow\left[\hat{X}^{\mu}, \hat{P}_{\nu}\right]=i \hbar \delta_{\nu}^{\mu} .
\end{equation}
The 16 independent variables of classical Fourier-phase space $x^{\mu}, p_{\nu}, k_{\rho}$, and $q^{\sigma}$ can be replaced by the 8 independent variables of quantum Fourier space (or quantum phase space) $\hat{X}^{\mu}$ and $\hat{P}_{\nu}$. This results in the the Heisenberg algebra with $\left[\hat{X}^{\mu}, \hat{X}^{\nu}\right]=\left[\hat{P}_{\mu}, \hat{P}_{\nu}\right]=0$.

The Heisenberg algebra can be found from the Koopman-von Neumann algebra by shifting coordinates \cite{Bondar}. However, the relationship between wavenumber and the variable conjugate to position was not mentioned, which overlooked the notion of $\hat{P}_{\mu}=\hbar \hat{K}_{\mu}$. An important property of quantum mechanics is that the eigenvalues of $\hat{X}^{\mu}$ should lead to coordinates of $\hat{x}^{\mu}$, assuming the quantum eigenvalues relate to a notion of classical reality. To understand what shifts in coordinates can be made, consider
\begin{eqnarray}
\hat{X}^{\mu} &=& a \hat{x}^{\mu}+b \hbar \hat{q}^{\mu}, \\
\hat{P}_{\nu} &=& c \hat{p}_{\nu}+d \hbar \hat{k}_{\nu},
\end{eqnarray}
where $a, b, c$, and $d$ are constants to be determined partially by the following constraint,
\begin{equation}
\left[\hat{X}^{\mu}, \hat{P}_{\nu}\right]=\left[a \hat{x}^{\mu}, c \hat{p}_{\nu}\right]+\left[a \hat{x}^{\mu}, d \hbar \hat{k}_{\nu}\right]+\left[b \hbar \hat{q}^{\mu}, c \hat{p}_{\nu}\right]+\left[b \hbar \hat{q}^{\mu}, d \hbar \hat{k}_{\nu}\right]=i \hbar(a d+b c) \quad \rightarrow \quad a d+b c=1 .
\end{equation}
The quantum theory should have operators $\hat{K}_{\rho}$ and $\hat{Q}^{\sigma}$ which satisfy the following relations,
\begin{equation}
\hat{P}_{\rho}=\hbar \hat{K}_{\rho}, \quad \hat{X}^{\sigma}=\hbar \hat{Q}^{\sigma},
\end{equation}
which results in
\begin{eqnarray}
\hat{K}_{\rho} &=& d \hat{k}_{\rho}+\frac{c}{\hbar} \hat{p}_{\rho}, \\
\hat{Q}^{\sigma} &=& b \hat{q}^{\sigma}+\frac{a}{\hbar} \hat{x}^{\sigma} .
\end{eqnarray}
In order to have a quantum theory whose operators $\hat{X}^{\mu}, \hat{P}_{\nu}, \hat{K}_{\rho}$, and $\hat{Q}^{\sigma}$ lead to the same eigenvalues as found in classical theory with $\hat{x}^{\mu}, \hat{p}_{\nu}, \hat{k}_{\rho}$, and $\hat{q}^{\sigma}$, the following operator relations are found when $\hat{x}^{\mu}=\hbar \hat{q}^{\mu}$ and $\hat{p}_{\nu}=\hbar \hat{k}_{\nu}$,
\begin{eqnarray}
\hat{X}^{\mu} &=& (a+b) \hat{x}^{\mu} \quad \rightarrow \quad a+b=1 \text {, } \\
\hat{P}_{\nu} &=& (c+d) \hat{p}_{\nu} \quad \rightarrow \quad c+d=1, \\
\hat{K}_{\rho} &=& (b+a) \hat{k}_{\rho} \quad \rightarrow \quad a+b=1, \\
\hat{Q}^{\sigma} &=& (d+c) \hat{q}^{\sigma} \quad \rightarrow \quad c+d=1 \text {. }
\end{eqnarray}
When the following relations are satisfied, the quantum theory satisfies $\hat{P}_{\nu}=\hbar \hat{K}_{\nu}=\hat{p}_{\nu}=\hbar \hat{k}_{\nu}$, etc. This assures that the quantum and classical Fourier phase space operators all agree such that the quantum eigenvalues correspond to classical variables as introduced in the Copenhagen interpretation. The solutions are not fully constrained, which allows for $b, c$, and $d$ to be found in terms of $a$,
\begin{equation}
b=1-a, \quad c=\frac{a-1}{2 a-1}, \quad d=\frac{a}{2 a-1} .
\end{equation}
Note that Ref.~\cite{Bondar} chose $a=1, b=1 / 2, c=1$, and $d=1 / 2$ (since their $\hat{\lambda}_{p}$ is our $-\hat{q}$). What this suggests is that the quantum theory in their formalism leads to quantum position eigenvalues that do not correspond to the position eigenvalues in the classical theory. For instance, if their $\hat{p}_{q}=\hbar \hat{k}_{q}$ and $\hat{p}=\hbar \hat{k}$, then $\hat{p}_{q}=\frac{3}{2} \hat{p}$. This was not considered problematic, as taking $\alpha$ (or $\hbar$) to zero leads to $\hat{p}_{q}=\hat{p}$. However, quantum theory leads to measurements of classical observables with real eigenvalues, so it seems to be an improvement to select only the classical formulations that admit a correspondence between the operators. While classical mechanics does not satisfy $\hat{p}=\hbar \hat{k}$, our assertion is that quantum theory must use operators $\hat{P}=\hbar \hat{K}$ that contain eigenvalues that match $\hat{P}=\hat{p}$ and $\hat{K}=\hat{k}$ in some classical framework, which implies that $\hat{p}=\hbar \hat{k}$ within the quantum theory.

\begin{figure}
\includegraphics[width=.8\textwidth]{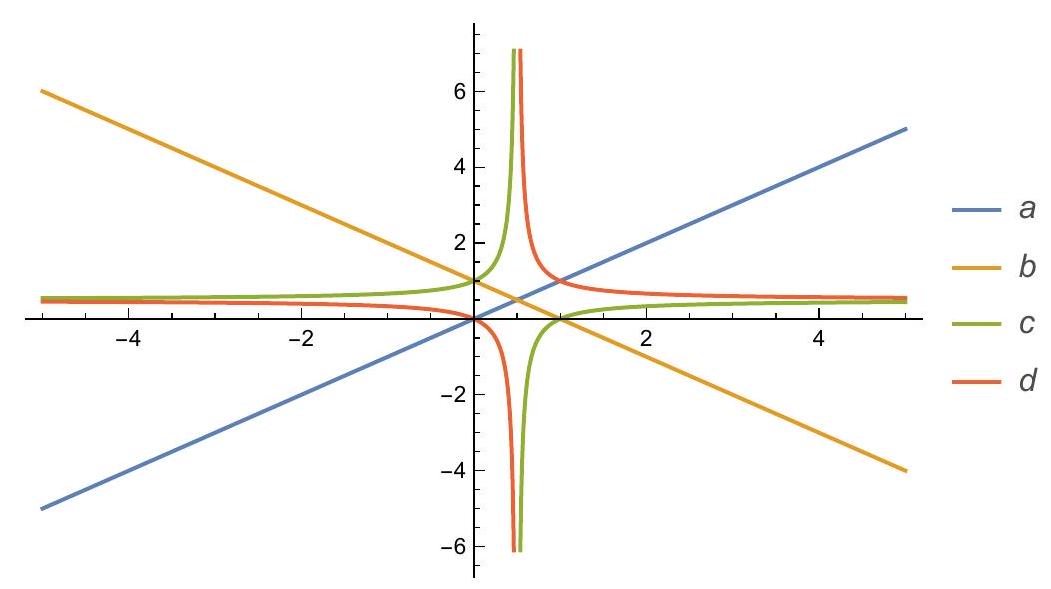}
\caption{The solution for coefficients $a, b, c$, and $d$ are shown as a function of $a$ such that both the classical and quantum phase space coordinates equal to Planck's constant times their corresponding wavenumbers.}
\label{solnFig}
\end{figure}

A few simple solutions for the coefficients include
\begin{equation}
\begin{array}{cccc}
\mbox{Solution 1: }a=1, & b=0, & c=0, & d=1, \\
\mbox{Solution 2: }a=2, & b=-1, & c=\frac{1}{3}, & d=\frac{2}{3}, \\
\mbox{Solution 3: }a=\frac{3}{2}, & b=-\frac{1}{2}, & c=\frac{1}{4}, & d=\frac{3}{4}, \\
\mbox{Solution 4: }a=\frac{1}{\sqrt{2}}, & b=1-\frac{1}{\sqrt{2}}, & c=-\frac{1}{\sqrt{2}}, & d=1+\frac{1}{\sqrt{2}},
\end{array} \label{solns}
\end{equation}
where all of these satisfy $a d+b c=1, a+b=1$, and $c+d=1$. The solutions for $a, b, c$, and $d$ are shown in Fig.~\eqref{solnFig}. The first solution above is clearly the simplest, as it leads to
\begin{eqnarray}
\hat{X}^{\mu} &=& \hat{x}^{\mu}, \\
\hat{P}_{\nu} &=& \hbar \hat{k}_{\nu}, \\
\hat{K}_{\rho} &=& \hat{k}_{\rho}, \\
\hat{Q}^{\sigma} &=& \frac{1}{\hbar} \hat{x}^{\sigma} .
\end{eqnarray}
This change of variables from Koopman-von Neumann classical mechanics to quantum mechanics is the simplest and most unambiguous, as it clearly states that the quantum momentum is equal to $\hbar$ times the classical wavenumber. For this solution, the classical momentum $\hat{p}_{\nu}$ and its conjugate momentum-wavenumber $\hat{q}^{\sigma}$ are not expressed in the quantum theory.

\section{Poly-Koopman-von Neumann mechanics as De Donder-Weyl theory}
\label{DDW}


Our next goal is to generalize KvN mechanics to relativistic field theory, which can be done in multiple ways, as there are multiple relativistic formulations of Hamiltonian dynamics for fields. While the Dirac or canonical Hamiltonian density is most popular, the De Donder-Weyl Hamiltonian density naturally contains the poly-momentum as found in the Euler-Lagrange equations for relativistic field theory, as shown in Table \eqref{LagrangeHamilton}. While both formulations have a momentum field, the Dirac momentum is typically called the canonical momentum, while the DDW momentum is covariant. 

Reconciling canonical and covariant formulations of quantum gravity is perceived as a challenge to this day \cite{Loll:2022ibq}. While formulations have been proposed by both string theorists and loop quantum gravity researchers, relatively little attention has been given to KvN \cite{Koopman,vonNeumann,Bondar,Morgan:2019azd,Morgan:2021jeh,Piasecki:2021keq,opmech2022} and DDW theory \cite{DeDonder1935,Weyl1934,Weyl1935,vonRieth}. While Witten et al.~have found canonical and covariant formulations with symplectic structure, no Hamiltonian density was found \cite{Witten1986, Zuckerman, Witten1987, Crnkovic1987, Crnkovic1988}. Researchers have recently explored Witten's use of covariant, canonical, and symplectic structures in a wide range of gravitational theories, often referred to as covariant phase space methods \cite{Margalef-Bentabol:2020teu,G:2021xvv,BarberoG:2021cei,G:2022ger}. 
DDW theory contains poly-symplectic/multisymplectic geometry, which contains a Hamiltonian density and a conjugate poly-momentum field. The covariant phase space methods refer to the conjugate poly-momentum, but the connection to DDW theory has been largely overlooked. 
Thiemann acknowledges that quantization of DDW theory is challenging and relatively unexplored \cite{Thiemann}, in reference to progress from Kantachikov on DDW quantization \cite{Kanatchikov:1996fx, Kanatchikov:1997pj, Kanatchikov:1997wp, Kanatchikov:1998yu, Kanatchikov:1998xz, Kanatchikov:1998vt, Kanatchikov:1999ut, Kanatchikov:2001uf, Kanatchikov:2002yh,Kanatchikov:2008he,Kanatchikov:2012gr, Kanatchikov:2013xmu, Kanatchikov:2015hna, Kanatchikov:2017zfe, Kanatchikov:2018uoy}.


A common approach to quantization is canonical quantization, which uses Dirac's $i\hbar$ prescription by converting classical Poisson brackets into quantum commutation relations. Groenewold's theorem demonstrates that such a quantization map does not formally exist in general, while deformation quantization considers a Moyal bracket as the appropriate quantum deformation of Poisson brackets. Geometric quantization includes prequantization, which introduces a prequantum Hilbert space with a mapping between classical Poisson brackets to quantum commutators. Kanatchikov's precanonical quantization is a type of geometric quantization from a DDW Hamiltonian, which studies Gerstenhaber brackets as a generalization of Poisson brackets \cite{Kanatchikov:2001uf}.

Kanatchikov has extensively explored DDW theory with a poly-symplectic form $\Omega$ \cite{Kanatchikov:1996fx, Kanatchikov:1997pj, Kanatchikov:1997wp, Kanatchikov:1998yu, Kanatchikov:1998xz, Kanatchikov:1998vt, Kanatchikov:1999ut,Kanatchikov:2001uf, Kanatchikov:2002yh, Kanatchikov:2008he, Kanatchikov:2012gr, Kanatchikov:2013xmu, Kanatchikov:2015hna, Kanatchikov:2017zfe, Kanatchikov:2018uoy}. 
His precanonical quantization extends the DDW Hamiltonian operator to act on a Clifford algebra-valued wavefunction with Dirac's gamma matrices giving $i\hbar \kappa \gamma^\mu \partial_\mu \Psi = \hat{H} \Psi$ as a generalized Schr\"odinger equation for $\Psi$ in a Clifford algebra \cite{Kanatchikov:1996fx}. While DDW theory directly applies to classical fields, Kanatchikov's exploration of precanonical quantization is largely based on Clifford-valued wavefunction, similar to how the Wheeler-DeWitt equation has a wavefunctional. While authors have explored DDW theory \cite{McLean, Helein, Forger, Nikolic, Riahi:2019tyb, Berra-Montiel:2017led, Berra-Montiel:2019ajm, Berra-Montiel:2021ypj}, most do not discuss quantization outside of Kanatchikov's work.  Kanatchikov's generalized Schr\"odinger equation may also be significant for Clifford relativity and applications to membranes \cite{Castro2010,Castro2014,Castro2014b,Castro2015,Castro2016,Castro2017,Castro2022}.

Below, we use poly-Poisson brackets previously explored by Kanatchikov \cite{Kanatchikov:1999ut} to obtain the equations of motion, which makes Dirac's quantization not possible and led Kanatchikov to consider the Gerstenhaber brackets. However, by introducing new poly-KvN algebras in addition to the poly-Poisson brackets, a transparent quantization of DDW theory is found by mapping commutators of classical operators to commutators of quantum operators. 
Both geometric quantization and KvN quantization have a notion of polarization, which relates to the mapping of $2d$ dimensions of phase space to $d$ dimensions of space or spacetime. The KvN quantization introduced by Bondar et al.~\cite{Bondar} is therefore similar to geometric quantization, except the quantization map most straightforwardly maps operators in a complex Hilbert space from a classical commutator algebra to a quantum commutator algebra. The nonrelativisti KvN algebra contains operators $\hat{x}^{i}$, $\hat{p}_{i}$, $\hat{k}_{i}$, and $\hat{q}^{i}$. We clarify how quantization of KvN mechanics relates to setting $\hat{p}_i = \hbar \hat{k}_i$ as operators. This quantization procedure is in this sense straightforward and intuitive.

\subsection{Classical field theory with De Donder-Weyl theory}

Classical non-relativistic Lagrangian formulations lead to Euler-Lagrange equations involving a time derivative $\frac{d}{dt}$. This naturally leads to a covariant momentum in terms of a velocity that takes a time derivative, giving a Hamiltonian. Relativistic field theory generalizes the Euler-Lagrange equations to include a partial derivative $\partial_\mu$. However, the so-called canonical Hamiltonian formulation from Dirac does not consider a poly-momentum from $\partial_\mu$ and instead resorts to the time derivative within some frame of reference, which breaks manifest Lorentz symmetry and is not covariant \cite{Dirac1950}. As we saw for the relativistic point particle, using the proper time with $\frac{d}{d\tau}$ instead of $\frac{d}{dt}$ also leads to a vanishing proper Hamiltonian. A natural framework for obtaining a manifestly covariant and canonical Hamiltonian density that generalizes the notion of the relativistic Euler-Lagrange equations is to introduce a poly-momentum that uses $\partial_\mu$ instead of $\frac{d}{dt}$ \cite{DeDonder1935,Weyl1934,Weyl1935}. Relativistic field theories motivate poly-symplectic geometry, since the Lagrangian density integrates over spacetime, not time. To summarize, consider the Euler-Lagrange equations in classical mechanics vs classical field theory with a Klein-Gordon scalar, which lead to the corresponding Hamiltonian formulations as shown in Table \eqref{LagrangeHamilton}.

\begin{table}
\begin{tabular}{c||c|c}
& Euler-Lagrange & Hamiltonian \\ \hline\hline
Newtonian particles & $\frac{d}{dt} \frac{d L}{d v^i} - \frac{dL}{dx^i} = 0$ & $H = v^ip_i - L$ \\ \hline
Relativistic fields & $\partial_\mu \frac{\partial \mathcal{L}}{\partial \nu_\mu^A} - \frac{\partial \mathcal{L}}{\partial \phi^A} = 0$ & $\mathcal{H} = \nu_\mu^A \pi^\mu_A - \mathcal{L}$
\end{tabular}
\caption{A comparison of the Euler-Lagrange equations with the Hamiltonian (density) for Newtonian mechanics and relativistic field theory is shown above. Since the Euler-Lagrange equations contain a partial derivative with respect to the poly-velocity $\nu_\mu = \partial_\mu \phi^A$, the De Donder-Weyl (DDW) Hamiltonian density $\mathcal{H}$ with poly-momentum $\pi^{\mu A}$ conjugate to any field $\phi^A$ is natural for a manifestly covariant and canonical Hamiltonian formulation.}
\label{LagrangeHamilton}
\end{table}

A manifest covariant and canonical formulation motivates the De Donder-Weyl Hamiltonian density $\mathcal{H}$ that is different than the canonical Hamiltonian density $\mathcal{H}_D$. The poly-velocity and poly-momentum densities of any field $\phi$ are given by
\begin{eqnarray}
\nu_{\mu} &=& \partial_{\mu} \phi \\
\pi^{\mu} &=& \frac{\partial \mathcal{L}}{\partial \partial_{\mu} \phi}
\end{eqnarray}
Since $\partial_{0}=\frac{1}{c} \frac{\partial}{\partial t}$, the standard canonical momentum density $\pi=\frac{\partial \mathcal{L}}{\partial \dot{\phi}}=\frac{1}{c} \pi^{0}$. 
Consider a different type of Poisson bracket that contains a covector index to express the poly-symplectic geometry,
\begin{equation}
\{A,B\}_\mu = \frac{\partial A}{\partial \phi(x)} \frac{\partial B}{\partial\pi^\mu(x)} - \frac{\partial A}{\partial \pi^\mu(x)}\frac{\partial B}{\partial \phi(x)} = \int d^4y \delta^{(4)}(x-y) \left[ \frac{\partial A}{\partial \phi(x)} \frac{\partial B}{\partial\pi^\mu(y)} - \frac{\partial A}{\partial \pi^\mu(y)}\frac{\partial B}{\partial \phi(x)}\right]. 
\end{equation}
The poly-symplectic structure of phase space over fields is characterized by this poly-Poisson bracket, since
\begin{eqnarray}
\{ \phi(x), \pi^\mu(x)\}_\nu &=& \delta^\mu_\nu, \\
\{\phi(x), \phi(x)\}_\mu &=& \{\pi^\mu(x), \pi^\nu(x)\}_\rho = 0.
\end{eqnarray}
Kanatchikov had previously discussed the same type of brackets in the language of differential forms \cite{Kanatchikov:1999ut}, which is similar to other authors use of poly-symplectic Poisson brackets \cite{Forger,Good,Tapia,Struckmeier,McClain, Blacker, Fernandes}.  
Other Poisson brackets over fields lead to Dirac delta functions over space \cite{Loll:2022ibq}. Our conventions for the poly-Poisson brackets allow for an easy identification of the De Donder-Weyl equations in a manner that mimics classical mechanics. 
Note that Dirac's $i\hbar$ prescription no longer holds, since the poly-symplectic geometry has a conjugate poly-momentum of a different rank than the original field. This does not obscure second quantization, especially once the generalization of the Koopman-von Neumann algebra is articulated. 

The De Donder-Weyl Hamiltonian density is
\begin{equation}
\mathcal{H}=\pi^{\mu} \nu_{\mu}-\mathcal{L} .
\end{equation} 
The De Donder-Weyl Hamiltonian is different than the canonical Hamiltonian introduced by Dirac, as it contains additional Legendre transformations for spatial components of the poly-momentum and poly-velocity. 
The canonical form of the action is therefore
\begin{equation}
S = \int d^4x \left( \nu_\mu \pi^\mu - \mathcal{H} \right).
\end{equation}
The variation of the action with respect to a classical orbit $\phi = \phi_{\textrm{cl}} + \delta \phi$ and $\pi^\mu = \pi^\mu_{\textrm{cl}} + \delta \pi^\mu$ leads to 
\begin{eqnarray}
\delta S &=& \int d^4x \left[\delta \nu_\mu \pi^\mu + \nu_\mu \delta\pi^\mu - \frac{\partial \mathcal{H}}{\partial\phi} \delta\phi - \frac{\partial \mathcal{H}}{\partial \pi^\mu} \delta \pi^\mu \right], \\ 
&=& \int d^4x \left[ \left(\nu_\mu - \frac{\partial\mathcal{H}}{\partial \pi^\mu} \right)\delta\pi^\mu - \left(\partial_\mu \pi^\mu + \frac{\partial\mathcal{H}}{\partial\phi} \right) \delta\phi + \partial_\mu \left( \pi^\mu \delta\phi\right) \right]. \nonumber
\end{eqnarray}
Assuming that boundary terms vanish, the De Donder-Weyl equations are found to give 
\begin{eqnarray}
\partial_\mu \phi(x) &=& \frac{\partial\mathcal{H}}{\partial \pi^\mu (x)} = \{\phi (x),\mathcal{H}\}_\mu, \\
\partial_\mu\pi^\mu(x) &=& - \frac{\partial\mathcal{H}}{\partial\phi (x)} = \frac{1}{d}\{\pi^\mu(x),\mathcal{H}\}_\mu ,
\end{eqnarray}
where $\frac{\partial \pi^\mu}{\partial \pi^\nu} = \delta^\mu_\nu$ and $\delta^\mu_\mu = d$ for $d$-dimensional spacetime.

The canonical Hamiltonian density allows for the derivation of field equations. 
Before discussing quantization, we demonstrate how derivations of classical equations of motion are simpler with the De Donder-Weyl Hamiltonian density in comparison to the canonical Hamiltonian density. The Hamilton-Jacobi theory for De Donder-Weyl theory for Klein-Gordon, Dirac, and gauge fields has been previously discussed \cite{vonRieth}. Kanatchikov has discussed a wide variety of field theories with De Donder-Weyl theory, including p-form electrodynamics, Yang-Mills theory, and general relativity to name a few \cite{Kanatchikov:1996fx, Kanatchikov:1997pj, Kanatchikov:1997wp, Kanatchikov:1998yu, Kanatchikov:1998xz, Kanatchikov:1998vt, Kanatchikov:2001uf, Kanatchikov:2002yh, Kanatchikov:2012gr, Kanatchikov:2013xmu, Kanatchikov:2015hna, Kanatchikov:2017zfe, Kanatchikov:2018uoy}. Next, we provide simple and straightforward derivations of the Klein-Gordon, Yang-Mills, and linearized gravity fields using standard notation familiar to quantum field theorists. Later, we show that KvN quantization can be applied to DDW theory after generalizing the KvN algebra to poly-symplectic fields. 

\subsubsection{Klein-Gordon scalar field}
The action of a real and massive Klein-Gordon scalar field $\phi(x)$ with an arbitrary potential $V(\phi)$ is
\begin{equation}
S_{K G}=\int d^{4} x\left(\frac{1}{2} \partial_{\mu} \phi \partial^{\mu} \phi-\frac{m^{2}c^2}{2\hbar^2} \phi^{2}-V(\phi)\right), 
\end{equation}
where the mostly-minus metric $\eta_{\mu \nu}$ is used. Throughout, $\hbar = c = 1$. The Klein-Gordon poly-velocity and poly-momentum are
\begin{equation}
\nu_{\mu}=\partial_{\mu} \phi, \quad \pi^{\mu}=\frac{\partial \mathcal{L}}{\partial \nu_{\mu}}=\partial^{\mu} \phi .
\end{equation}
The De Donder-Weyl Hamiltonian density for the Klein-Gordon theory is 
\begin{equation}
\mathcal{H}=v_{\mu} \pi^{\mu}-\mathcal{L}=\frac{1}{2} \pi^{\mu} \pi_{\mu}+\frac{m^{2}}{2} \phi^{2}+V(\phi)
\end{equation}
The De Donder-Weyl equations give
\begin{eqnarray}
\partial_{\mu} \phi &=& \frac{\partial \mathcal{H}}{\partial \pi^{\mu}}=\pi_{\mu}, \\
\partial_{\mu} \pi^{\mu} &=& -\frac{\partial \mathcal{H}}{\partial \phi}=-m^{2} \phi-\frac{\partial V}{\partial \phi}
\end{eqnarray}
Plugging the first equation into the second equation gives the Klein-Gordon equation,
\begin{equation}
\left(\partial_{\mu} \partial^{\mu}+m^{2}\right) \phi=-\frac{\partial V}{\partial \phi} .
\end{equation}
Since this Hamiltonian formulation is manifestly relativistic, it is more straightforward than the canonical Hamiltonian approach. The tradeoff was to introduce a poly-momentum vector field that is conjugate to the scalar field, but the poly-velocity is already in the standard Euler-Lagrange equations for fields. 

Since it will be instructive for a later demonstration, we also consider a complex Klein-Gordon field with the following action,
\begin{equation}
S = \int d^4x \left( \partial_\mu \phi \partial^\mu \phi^* - m^2|\phi|^2 - V(\phi,\phi^*)\right).
\end{equation}
Since $\phi = \phi_1 + i \phi_2$, the poly-Poisson brackets must sum over both $\phi$ and $\phi^*$. The following poly-velocities and poly-momenta are found,
\begin{eqnarray}
\nu_\mu = \partial_\mu \phi, &\qquad& \pi^\mu = \frac{\partial\mathcal{L}}{\partial \nu_\mu} = \partial^\mu \phi^*, \nonumber \\ 
\nu^*_\mu = \partial_\mu \phi^*, &\qquad& \pi^{*\mu} = \frac{\partial\mathcal{L}}{\partial \nu_\mu^*} = \partial^\mu \phi, 
\end{eqnarray}
where $\nu_\mu^*$ is defined as the poly-velocity of $\phi^*$, not immediately as the complex conjugate of $\nu_\mu$. 
The poly-Poisson bracket must consider both $\phi$ and $\phi^*$,
\begin{equation}
\{ A, B\}_\mu = \frac{\partial A}{\partial \phi} \frac{\partial B}{\partial\pi^\mu} - \frac{\partial A}{\partial \pi^\mu} \frac{\partial B}{\partial \phi} + \frac{\partial A}{\partial \phi^*} \frac{\partial B}{\partial\pi^{*\mu}} - \frac{\partial A}{\partial \pi^{*\mu}} \frac{\partial B}{\partial \phi^*} .
\end{equation}
For later convenience, the De Donder-Weyl Hamiltonian density is given by 
\begin{equation}
\mathcal{H} = \nu_\mu \pi^\mu + \nu_\mu^* \pi^{*\mu} - \mathcal{L} = \pi^\mu \pi_\mu^* + m^2 |\phi|^2 + V(\phi,\phi^*).
\end{equation}

\subsubsection{Yang-Mills spin-1 field}
Next, the DDW Hamiltonian for Yang-Mills theory is derived. 
The spin-1 Yang-Mills field has the following action,
\begin{equation}
S=\int d^{4} x\left(-\frac{1}{4} F_{\mu \nu}^{A} F^{\mu \nu A}-J^{\mu A} A_{\mu}^{A}\right) ,
\end{equation}
where the Yang-Mills field strength $F_{\mu\nu}^A$ is found in terms of the gauge potential $A_\mu^A$,
\begin{equation}
F_{\mu\nu}^A = \partial_\mu A_\nu^A - \partial_\nu A_\mu^A - gf^{ABC} A_\mu^b A_\nu^c. 
\end{equation}
The Yang-Mills poly-velocity and poly-momentum are
\begin{equation}
\nu_{\mu \nu}{ }^{A}=\partial_{\mu} A_{\nu}^{A}, \quad \pi^{\mu \nu A}=-F^{\mu \nu A} .
\end{equation}

The De Donder-Weyl Hamiltonian density for Yang-Mills theory is
\begin{equation}
\mathcal{H}=-\frac{1}{4} \pi_{\mu \nu}^{A} \pi^{\mu \nu A}+\frac{g}{2} \pi^{\mu \nu A} f^{A B C} A_{\mu}^{B} A_{\nu}^{C}+J^{\mu A} A_{\mu}^{A} .
\end{equation}
The De Donder-Weyl equations give the equations of motion for Yang-Mills theory when asserting that $\pi^{\mu \nu A}$ is antisymmetric,
\begin{eqnarray}
\frac{1}{2}\left(\partial_{\mu} A_{\nu}^{A}-\partial_{\nu} A_{\mu}^{A}\right) &=& \frac{\partial \mathcal{H}}{\partial \pi^{\mu \nu A}}=-\frac{1}{2} \pi_{\mu \nu}{ }^{A}+\frac{g}{2} f^{A B C} A_{\mu}^{B} A_{\nu}^{C} \\
\partial_{\mu} \pi^{\mu \nu A} &=& -\frac{\partial \mathcal{H}}{\partial A_{\nu}^{A}}= -\frac{g}{2} \pi^{\nu \mu B} f^{B A C} A_{\mu}^{C}-\frac{g}{2} \pi^{\mu \nu C} f^{C B A} A_{\mu}^{B}-J^{\nu A}=g f^{A B C} A_{\mu}^B \pi^{\mu \nu C}-J^{\nu A}
\end{eqnarray}
The first equation leads to $\pi^{\mu \nu A}=-F^{\mu \nu A}$, which when plugging this into the second equation gives the Yang-Mills equations of motion,
\begin{equation}
\partial_{\mu} F^{\mu \nu A} = J^{\nu A} + gf^{ABC}A_\mu^B F^{\mu\nu C} = j^{\nu A}, 
\end{equation}
where $j^{\nu A}$ is the gauge-dependent current. The equations of motion are also expressed as $D_\mu F^{\mu\nu A} = J^{\nu A}$, where $D_\mu$ is the covariant derivative. Choosing the Abelian gauge group $U(1)$ would lead to Maxwell's electrodynamics. This completes the derivation of the equations of motion for all of the field content in the standard model within De Donder-Weyl theory.

Since the gauge-dependent Yang-Mills current will be inspirational for generalizing the Liouville equation in the relativistic setting, consider how charge conservation can be stated with the covariant or partial derivative,
\begin{equation}
D_\mu J^{\mu A} =\partial_\mu J^{\mu A} - gf^{ABC} A_\mu^B J^{\mu C}= 0, \qquad  \partial_\mu j^{\mu A} = 0,
\end{equation}
where $J^{\mu A}$ is the gauge-covariant current, while $j^{\mu A}$ is the conserved Noether current. Later, the poly-Liouville equation for probability 4-current conservation of relativistic fields will be derived in analogy with $D_\mu J^{\mu A} = 0$. 

\subsubsection{Linearized gravity with spin-2 field}


In linearized gravity, the spacetime metric $g_{\mu\nu}$ is perturbed with respect to a background Minkowski metric $\eta_{\mu\nu}$,
\begin{eqnarray}
g_{\mu\nu} &=& \eta_{\mu\nu} + \kappa h_{\mu\nu}, \\ 
g^{\mu\nu} &=& \eta^{\mu\nu} - \kappa h^{\mu\nu} + \mathcal{O}(\kappa^2),
\end{eqnarray}
where $h_{\mu\nu}(x)$ is the linearized gravitational field and $\kappa = \sqrt{\frac{32 \pi G}{c^3}}$. The linearized gravity action is 
\begin{equation}
S = \int d^4x \left( \frac{1}{2}\partial^\rho h^{\mu\nu} \partial_\rho h_{\mu\nu} - \partial_\mu h^{\mu\rho} \partial^\nu h_{\nu\rho} + \partial^\mu h_{\mu\nu}\partial^\nu h-\frac{1}{2}\partial_\mu h\partial^\mu h\right),
\end{equation}
where $h = h_{\mu\nu}\eta^{\mu\nu}$. The trace-reversed metric $\bar{h}_{\mu\nu} = h_{\mu\nu} -\frac{1}{2} \eta_{\mu\nu} h$ is the field mapped from the Yang-Mills gauge field with respect to the radiative double copy \cite{Chester:2017vcz}. This allows for a simplification in the action and the field equations as well, since
\begin{equation}
S = \int d^4x \left( \frac{1}{2} \partial^\rho \bar{h}^{\mu\nu}\partial_\rho \bar{h}_{\mu\nu} - \frac{1}{4}\partial_\mu \bar{h}\partial^\mu \bar{h} \right). 
\end{equation}
Plugging $\bar{h}_{\mu\nu}$ in terms of $h_{\mu\nu}$ of the action above leads to
\begin{equation}
S = \int d^4x \left( \frac{1}{2} \partial^\rho h^{\mu\nu}\partial_\rho h_{\mu\nu} - \frac{1}{4}\partial_\mu h\partial^\mu h \right). 
\end{equation}

Consider the poly-velocity and poly-momentum of the trace-reversed metric,
\begin{eqnarray}
\bar{\nu}_{\rho\mu\nu} &=& \partial_\rho \bar{h}_{\mu\nu}, \\ 
\bar{\pi}^{\rho\mu\nu} &=& \frac{\partial\mathcal{L}}{\partial \bar{\nu}_{\rho\mu\nu}} = \partial^\rho \bar{h}^{\mu\nu} - \frac{1}{2}\eta^{\mu\nu}\partial^\rho \bar{h} = \partial^\rho h^{\mu\nu},
\end{eqnarray}
where the poly-momentum of the trace-reversed metric is the partial derivative of the metric. The De Donder-Weyl Hamiltonian density for linearized gravity is
\begin{eqnarray}
\mathcal{H} &=& \bar{\nu}_{\rho\mu\nu} \bar{\pi}^{\rho\mu\nu} -\mathcal{L} = \frac{1}{2}\partial^\rho \bar{h}^{\mu\nu} \partial_\rho \bar{h}_{\mu\nu} - \frac{1}{4} \partial_\rho \bar{h}\partial^\rho\bar{h} = \mathcal{L} \nonumber \\
&=& \frac{1}{2} \bar{\pi}^{\rho\mu\nu}\bar{\pi}_{\rho\mu\nu} - \bar{\pi}_{\mu}^{\,\,\,\mu\rho}\bar{\pi}^{\nu}_{\,\,\,\nu\rho} + \bar{\pi}^{\mu}_{\,\,\,\mu \nu} \eta^{\alpha\beta} \bar{\pi}^\nu_{\,\,\,\alpha\beta} - \frac{1}{2} \eta^{\alpha\beta}\eta^{\rho\sigma}\bar{\pi}_{\mu\alpha\beta}\bar{\pi}^\mu_{\,\,\,\rho\sigma},
\end{eqnarray}
where the massless linearized graviton's De Donder-Weyl Hamiltonian density is the same as the Lagrangian density, which is encouraging, as there is only gravitational kinetic energy present. 
The De Donder-Weyl equations give 
\begin{eqnarray}
\partial_\mu \bar{h}_{\alpha\beta} &=& \frac{\partial \mathcal{H}}{\partial \bar{\pi}^{\mu\alpha\beta}} = \bar{\pi}_{\mu\alpha\beta} - 2\eta_{\mu\alpha} \bar{\pi}^\nu_{\,\,\,\nu\beta} + \eta_{\mu\alpha}\eta^{\rho\sigma}\bar{\pi}_{\beta\rho\sigma} + \bar{\pi}^\rho_{\,\,\,\rho\mu}\eta_{\alpha\beta} - \eta_{\alpha\beta}\eta^{\rho\sigma}\bar{\pi}_{\mu\rho\sigma}, \label{linGrav1} \\ 
\partial_\mu \bar{\pi}^{\mu\alpha\beta} &=& - \frac{\partial \mathcal{H}}{\partial \bar{h}_{\alpha\beta}} = 0.
\end{eqnarray}
The equation of motion can be found by solving the first equation for $\pi^{\mu\alpha\beta}$ and plugging into the second equation. 

The first equation of motion differs from the assignment of the poly-momentum specifically unless the De Donder gauge condition is taken,
\begin{equation}
\partial_\mu \bar{h}^{\mu\nu} = 0.
\end{equation}
This ``gauge condition'' is sometimes called the harmonic, Hilbert, Lorenz, Lorentz, or Fock gauge. 
Specifically when this gauge is taken, then the definition of the poly-momentum found from the Lagrangian density can be found, which leads to the following vacuum equations of motion,
\begin{equation}
\partial_\mu \partial^\mu h^{\alpha\beta} = 0.
\end{equation}

Since the stress-energy-momentum tensor $T^{\mu\nu}$ is given by
\begin{equation}
T^{\mu\nu} = -\frac{2c}{\sqrt{-g}} \frac{\partial \sqrt{-g} \mathcal{L}_M}{\partial g_{\mu\nu}} = -2c \frac{\partial \mathcal{L}_M}{\partial g_{\mu\nu}} + c\mathcal{L} g^{\mu\nu},
\end{equation}
it is understood that $\kappa \bar{h}_{\mu\nu}$ will couple to $\bar{T}^{\mu\nu}$ at lowest order. This implies that the matter Lagrangian is
\begin{equation}
S_M = -\frac{\kappa}{2c} \int d^4x \sqrt{-g} \bar{T}^{\mu\nu}\bar{h}_{\mu\nu}.
\end{equation}
To derive the De Donder-Weyl Hamiltonian for linearized gravity in another manner, consider the solution for $\partial_\mu \bar{h}_{\alpha\beta}$ in terms of $\pi_{\mu\alpha\beta}$ from Eq.~\eqref{linGrav1},
\begin{equation}
\bar{\nu}_{\mu\alpha\beta}= \partial_\mu \bar{h}_{\alpha\beta} = \bar{\pi}_{\mu\alpha\beta} - \frac{1}{2}\eta_{\alpha\beta} \bar{\pi}_\mu \equiv \pi_{\mu\alpha\beta},
\end{equation}
A reciprocal relationship is found between the trace-reversed poly-velocity $\bar{\nu}_{\mu\alpha\beta}$ with the trace reverse of the trace-reversed poly-momentum $\pi_{\mu\alpha\beta}$. This allows for the De Donder-Weyl Hamiltonian density to be written as
\begin{equation}
\mathcal{H} = \frac{1}{2}\pi_{\mu\alpha\beta} \pi^{\mu\alpha\beta} - \frac{1}{4} \pi_\mu \pi^\mu+\frac{8\pi G}{c^4}\kappa \bar{T}^{\mu\nu} \bar{h}_{\mu\nu} = \frac{1}{2}\bar{\pi}_{\mu\alpha\beta} \bar{\pi}^{\mu\alpha\beta} - \frac{1}{4} \bar{\pi}_\mu \bar{\pi}^\mu+\frac{8\pi G}{c^4}\kappa \bar{T}^{\mu\nu} \bar{h}_{\mu\nu},
\end{equation}
where $\bar{\pi}_\mu = -\pi_\mu = \partial_\mu h = -\partial_\mu \bar{h}$. 
For this Hamiltonian density, the De Donder gauge was not explicitly solved for, but the reciprocal relationship of the polymomenta and the polyvelocities implies the De Donder gauge, as the De Donder-Weyl-Hamilton equations from the Hamiltonian density above lead to
\begin{eqnarray}
\partial_\mu \bar{h}_{\alpha\beta} &=& \frac{\partial \mathcal{H}}{\partial \bar{\pi}^{\mu\alpha\beta}} = \bar{\pi}_{\mu\alpha\beta} - \frac{1}{2}\eta_{\alpha\beta}\bar{\pi}_\mu = \pi_{\mu\alpha\beta}, \\
\partial_\mu \bar{\pi}^{\mu\alpha\beta} &=& - \frac{\partial\mathcal{H}}{\partial\bar{h}_{\alpha\beta}} = -\kappa \frac{8\pi G}{c^4} \bar{T}^{\alpha\beta}.
\end{eqnarray}
The trace-reversal of the second equation above after plugging in the first equation leads to the equations of motion for linearized gravity sourced by matter in the De Donder gauge,
\begin{equation}
\partial_\mu \pi^{\mu\alpha\beta} = \partial_\mu\partial^\mu \bar{h}^{\alpha\beta} = -\frac{8\pi G\kappa}{c^4} T^{\alpha\beta}. 
\end{equation}
The De Donder gauge is therefore a type of Hamiltonian constraint. Just as Dirac's canonical Hamiltonian formulation of general relativity reduces the metricial degrees of freedom from 10 to 6, so does imposing the De Donder gauge condition \cite{Dirac1958}. For general relativity, $\bar{h}_{\mu\nu}$ would be generalized to $\sqrt{-g}g_{\mu\nu}$, which has been discussed previously \cite{Horava1991,Cheung:2016say}

\subsection{Poly-Koopman-von Neumann mechanics as De Donder-Weyl theory}

Next, we generalize relativistic Koopman-von Neumann mechanics to classical field theory with poly-symplectic fields and recover De Donder-Weyl theory.  First, Koopman-von Neumann mechanics must be generalized to work on Fock spaces to describe classical fields. Second, a poly-Liouville operator $\hat{\mathfrak{L}}_\mu$ is constructed, which leads to a poly-Liouville equation for probability 4-current conservation. The classical analogue of a poly-Schr\"odinger equation with $\partial_\mu$ instead of $\frac{\partial}{\partial t}$ is also found, which corresponds to the first De Donder-Weyl equation, while the second De Donder-Weyl equation of motion is interpreted as a poly-Schr\"odinger equation for the conjugate poly-momentum. Finally, we formulate two types of poly-Koopman-von Neumann algebras with covariant and canonical commutation relations over a Fourier-phase space for classical and 1st quantized fields with poly-symplectic geometry. 


\subsubsection{Sketch of assumptions for axiomatic interacting field theories}


A Fock space is defined as the tensor product of $N$-particle Hilbert spaces. In quantum field theory, a non-interacting bare vacuum $|0\rangle$ is typically presumed to be embedded in Minkowski space with the Wightman or Haag-Kastler axioms \cite{Wightman,HaagKastler}. Hall and Wightman concluded from Haag's theorem that there is not a single Hilbert space representation for the free and interacting fields \cite{HallWightman}. We embrace this notion by considering a dynamical discrete spectrum of the renormalized (physical) vacuum state $|\Omega\rangle$ that is uniquely determined by the phase space geometry, which is not found in $|0\rangle$. The renormalized vacuum $|\Omega\rangle$ need not have the same spectrum as $|0\rangle$, since the setting of the energy scale relates to the types of measurements that can be performed. We suppose that algebraic quantum field theory can be constructed via operator algebras that need not be related to Minkowski space in the interacting field theory. From a philosophical perspective, QFTs should include gravity, since all fields carry energy and momentum as a source of gravitation. 

While quantum mechanics contains a measurement problem, quantum field theory does not in the same sense, as the renormalized measurement apparatus $\langle\Omega|$ contains additional information not found in the non-interacting bare template of abstract measurements $\langle 0|$. The true state of measurement is defined by a destruction operator acting on a renormalized vacuum $\langle\chi|=\langle\Omega| \hat{a}$, which denotes a notion of measurement via absorption of quantum information. Quantum mechanics has no destruction operators without Fock spaces. 

Since Fock spaces are used in field theory rather than Hilbert spaces, the wavefunction must be replaced with raising and lowering operators associated with quantum fields. If a field $\phi$ has $a^{\dagger}$ and $a$ as creation and annihilation operators that act on a renormalized vacuum state $|\Omega\rangle$, then projection operators are replaced via $\left|s^{\prime}\right\rangle\langle s| \rightarrow a_{s^{\prime}}^{\dagger} a_{s}$. Without rigorously developing axiomatic interacting QFT, the progression from wavefunctions $\psi(x)$ to (classical or quantum) fields $\phi(x)$ is aided by the Dirac bra-ket notation,
\begin{equation}
\hat{\mathcal{O}} \psi(x) \equiv\langle x|\hat{\mathcal{O}}| \psi\rangle \rightarrow \mathcal{O} \phi(x)\left\langle x\left|\hat{\mathcal{O}} a^{\dagger}\right| \Omega\right\rangle=\langle x|\hat{\mathcal{O}}| \phi\rangle,
\end{equation}
where a single-particle state $|\phi\rangle=a^{\dagger}|\Omega\rangle$ was constructed above. Without loss of generality, $|\phi\rangle$ can contain an arbitrary number of creation operators acting on $|\Omega\rangle$. In this manner, the classical Koopman-von Neumann wavefunction can be appropriately generalized to classical Koopman-von Neumann fields, including the relativistic Klein-Gordon field.

\subsubsection{Poly-Koopman-von Neumann mechanics}
\label{polyKvN}


Manifestly covariant and canonical evolution can be found from the De Donder-Weyl equations, which motivates the introduction of a covector poly-Liouville operator in poly-Koopman-von Neumann mechanics. The poly-Liouville operator is introduced as
\begin{equation}
i \hat{\mathfrak{L}}_{\mu} A=\{A,\mathcal{H}\}_{\mu}=\sum_{\phi}\left(\frac{\partial \mathcal{H}}{\partial \pi^\mu}\frac{\partial }{\partial \phi} - \frac{\partial \mathcal{H}}{\partial \phi} \frac{\partial }{\partial \pi^\mu}\right) A. \label{polyPoisson}
\end{equation}
If a wavefunction was evolved, then the classical poly-Koopman-von Neumann equation would be $i \partial_{\mu} \psi=\hat{\mathfrak{L}}_{\mu} \psi$ for a wavefunction $\psi$. However, we desire fields, not wavefunctions. As mentioned, since Fock spaces are constructed from Hilbert spaces, it is anticipated that a similar equation should hold for classical fields over Fock space. As such, the poly-Koopman-von Neumann equation should be generalized to apply the Liouville operator to the poly-symplectic fields, rather than a single wavefunction, which leads to two DDW equations instead of one. 

Consider the poly-Liouville operator acting on the poly-symplectic fields $\phi$ and $\pi^\mu$. The poly-Koopman-von Neumann equations for fields are defined as the poly-Liouville operator acting on classical fields of poly-symplectic phase space, which are found by applying the De Donder-Weyl equations, 
\begin{eqnarray}
i\hat{\mathfrak{L}}_\mu \phi &=& \{\phi, \mathcal{H}\}_\mu = \frac{\partial \mathcal{H}}{\partial \pi^\mu} =  \partial_\mu \phi, \\ 
i\hat{\mathfrak{L}}_\mu \pi^\nu &=& \{\pi^\nu, \mathcal{H}\}_\mu = - \frac{\partial \mathcal{H}}{\partial \phi} \frac{\partial \pi^\nu}{\partial \pi^\mu} = - \frac{\partial\mathcal{H}}{\partial \phi} \delta_\mu^\nu = \partial_\rho \pi^\rho\delta_\mu^\nu.
\end{eqnarray}
These poly-Koopman-von Neumann equations replace the classical wavefunction $\psi(x,p)$ with the classical fields $\phi$ and $\pi^\mu$. The poly-Liouville operator $\hat{\mathfrak{L}}_\mu$ acting on $\pi^\nu$ does not simply take the partial derivative $i\partial_\mu$, which is different than how the Liouville operator gives time evolution of the Koopman-von Neumann wavefunction similar to the Hamiltonian operator for the quantum wavefunction. Nevertheless, $i\partial_\mu \pi^\mu$ gives De Donder-Weyl equations, 
\begin{eqnarray}
i\partial_\mu \phi &=& -\hat{\mathfrak{L}}_\mu \phi = i\left\{\phi,\mathcal{H}\right\}_\mu = i \frac{\partial \mathcal{H}}{\partial \pi^\mu} , \\ 
i\partial_\mu \pi^\mu &=& -\frac{1}{d} \hat{\mathfrak{L}}_\mu \pi^\mu = \frac{i}{d}\left\{\pi^\mu, \mathcal{H}\right\}_\mu = - i\frac{\partial\mathcal{H}}{\partial \phi}.
\end{eqnarray}
This demonstrates for the first time that the De Donder-Weyl equations can be rewritten with a poly-Liouville operator, since $\{\pi^\mu,\mathcal{H}\}_\mu = -i\hat{\mathfrak{L}}_\mu \pi^\mu$. 

To clarify, Kanatchikov extends the Wheeler-De Witt equation to arbitrary DDW field theories, which is one type of generalized Schr\"odinger equation \cite{Kanatchikov:1998vt,Kanatchikov:2001uf}, while here, we simply point out that the first De Donder-Weyl equation is a generalization of the Scr\"odinger equation in another sense, as a relativistic field can be seen as a generalization of a wavefunction. In this sense, Kanatchikov's contributions to DDW theory are much deeper and significant than our somewhat trivial recovery of the DDW equations from poly-KvN mechanics. 

The relationship of the Liouville equation and Hamilton's equations motivates a poly-Liouville equation from De Donder-Weyl equations, which replaces the partial time derivative with the partial derivative of spacetime $\partial_\mu$. This implies that the poly-Liouville equation should involve a probability 4-current $J^\mu(\phi(x),\pi(x))$, rather than the probabilty density $\rho$. 
The probability 3-current of a wavefunction generalized to a 4-current leads to the following probability density for $\phi$ in quantum field theory,
\begin{equation}
J^\mu_{QFT} = \frac{i\hbar}{2m}\left( \phi^* \partial^\mu \phi - (\partial^\mu \phi^*)\phi\right).
\end{equation}
Classical theory does not have $\hbar$. Also, massless Klein-Gordon fields can be studied. 
Taking inspiration from above, consider the following canonical classical Klein-Gordon probability current for a complex scalar field,
\begin{equation}
J^\mu(\phi(x),\pi(x)) = \frac{i}{2}\left( \phi^* \pi^{*\mu} - \pi^{\mu} \phi\right).
\end{equation}



While the Liouville equation typically implements time evolution of global phase space with Poisson brackets of a Hamiltonian, the poly-Liouville equation implements local spacetime evolution with a Hamiltonian density. This analogy implies for the existence of a total spacetime derivative $\frac{d}{dx^\mu}$, which contains the partial spacetime derivative plus partial derivatives with respect to the fields and their conjugate polymomenta. It turns out to be trivial for real Klein-Gordon fields $\phi$, as $\partial_\mu J^\mu = 0$ and the Liouville operator on $J^\mu$ would vanish. For complex Klein-Gordon fields, the poly-Liouville equation can be derived by considering the following, 
\begin{eqnarray}
\partial_\mu J^\mu &= & \frac{i}{2} \left( \partial_\mu \phi^* \pi^{*\mu} + \phi^*\partial_\mu \pi^{*\mu} - \partial_\mu \pi^{\mu} - \pi^{\mu}\partial_\mu \phi \right)  = \frac{i}{2}\left( \phi^* \partial_\mu \pi^{*\mu} - \partial_\mu \pi^{\mu} \phi \right), \\ 
i\hat{\mathfrak{L}}_\mu J^\mu &=& \{J^\mu ,\mathcal{H} \}_{\mu} = \frac{\partial J^\mu}{\partial \phi}\frac{\partial \mathcal{H}}{\partial \pi^\mu} - \frac{\partial J^\mu}{\partial \pi^\mu}\frac{\partial \mathcal{H}}{\partial \phi} + \frac{\partial J^\mu}{\partial \phi^*}\frac{\partial \mathcal{H}}{\partial \pi^{*\mu}} - \frac{\partial J^\mu}{\partial \pi^{*\mu}}\frac{\partial \mathcal{H}}{\partial \phi^*} \\
&=& \frac{i d}{2}\left(\phi^*\partial_\rho \pi^{*\rho} - \phi \partial_\rho \pi^\rho\right).
\end{eqnarray}
Putting these two terms together, the poly-Liouville equation for arbitrary field theories is found to be
\begin{equation}
\frac{dJ^\mu}{dx^\mu} \equiv \partial_\mu J^\mu -\frac{i}{d} \hat{\mathfrak{L}}_\mu J^\mu = \partial_\mu J^\mu -\frac{1}{d}\{J^\mu,\mathcal{H} \}_\mu = \partial_\mu J^\mu -\frac{1}{d} \sum_i \partial_\mu \phi_i \frac{\partial J^\mu}{\partial\phi_i} + \partial_\nu \pi_i^\nu \frac{\partial J^\mu}{\partial \pi^\mu_i} = 0,
\end{equation}
where the complex Klein-Gordon scalar field theory has $\phi_i = (\phi, \phi^*)$.
The poly-symplectic geometry leads to a contraction between $\partial_\nu \pi^\nu$ in the last term above. Curiously, if $d_\mu \equiv \frac{d}{d x^\mu}$ is thought of as a covariant derivative, the poly-Liouville operator $\hat{\mathfrak{L}}_\mu$ is like a gauge field. 

The poly-Liouville equation describes the conservation of a 4-current density, whose structure is more similar to the second De Donder-Weyl equation. In this sense, the equations for the 4-current and the conjugate poly-momentum are similar, while the non-relativistic Liouville equation is more similar to the first De Donder-Weyl equation. The poly-Liouville equation acting on classical fields cannot be found by applying the chain rule due to the structure of poly-symplectic geometry. While this equation does give a partial derivative that includes a time derivative as used in time evolution, the poly-Liouville operator acting on $\phi$ gives $\pi^\mu$ rather than the evolution of $\phi$, as the poly-Liouville operator acts on $\pi^\mu$ to give the equations of motion for $\phi$ after plugging in the first equation. In summary, poly-KvN mechanics is equivalent to DDW theory. 




\subsubsection{0th and 1st quantized poly-Koopman-von Neumann algebras}
\label{classical-fields}


Finally, we consider poly-Koopman-von Neumann algebras that generalize the Koopman-von Neumann algebra. Since phase space coordinates are replaced with poly-symplectic fields, Fourier-conjugate fields $\kappa$ and $\xi^\mu$ are introduced,
\begin{equation}
\begin{array}{ccccccc}
x^\mu & & & \rightarrow & \phi  & & \\ 
k_\nu &=& -i\frac{\partial}{\partial x^\nu} & \rightarrow & \kappa &\propto& -i\frac{\partial}{\partial \phi} \\ 
p_\rho & & & \rightarrow & \pi^\rho & & \\
q^\sigma &=& i\frac{\partial}{\partial p_\sigma} & \rightarrow & \xi_\sigma &\propto& i\frac{\partial}{\partial \pi^\sigma}. 
\end{array}
\label{1st2ndFPS}
\end{equation}
While it is common to derive the classical equations of motion for the Klein-Gordon equation in terms of a scalar field $\phi(x)$, the poly-Koopman-von Neumann approach to field theory leads to a different type of classical field, as it may depend both on position and momentum. Since the poly-symplectic fields are generalizations of the wavefunction, the fully classical poly-Koopman-von Neumann fields should depend on position and momentum such as $\phi(x,p)$ or $\bar{\phi}(k,q)$ as Fourier-phase-space conjugates. First quantization is realized as quantizing phase space coordinates with $p_\mu = \hbar k_\mu$ without quantizing fields, which leads to $\phi(x)$ and $\phi(p)$ as Fourier conjugates. Second quantization applies quantization of the fields. As such, classical field theory admits 0th quantized (classical) and 1st quantized poly-Koopman-von Neumann algebras.

It is also worth clarifying that the study of KvN fields introduces the concept of 1st and 2nd Fourier conjugation. At the classical phase space level, a field $\phi(x,p)$ admits a 1st Fourier transformation to give $\tilde{\phi}(k,q)$. Half-Fourier transformations giving $\tilde{\phi}(k,p)$ or $\tilde{\phi}(x,q)$ can be found. In another sense, a functional such as $\mathcal{H}(\phi,\pi)$ may admit a 2nd Fourier transformation, leading to $\tilde{\mathcal{H}}(\kappa,\xi)$. In general, $\kappa(k,q)$ is not the 1st Fourier transform of $\phi(x,p)$, but rather $\kappa(x,p)$ is the 2nd Fourier conjugate field to $\phi(x,p)$. This matches how $x^\mu$ and $p_\nu$ denote coordinates for (1st) phase space, while the covariant phase space description refers to phase space over fields (2nd phase space). Kvn fields require the articulation of of 1st and 2nd Fourier-phase space over coordinates and fields as shown in Eq.~\eqref{1st2ndFPS}. Throughout, we refer to 1st quantization as the reduction of 1st Fourier-phase space, while 2nd quantization is the reduction of 2nd Fourier-phase space.

Overall, the discussion of generalizing Koopman-von Neumann mechanics has been rather limited. Gozzi and Reuter had previously considered the notion of classical path integrals as a counterpart to Koopman-von Neumann mechanics \cite{Gozzi}, which has been more recently reviewed by Piasecki \cite{Piasecki:2021keq}. However, Gozzi  and Reuter focused more on adding BRST ghosts to classical fields, rather than the 2nd Fourier conjugate fields such as $\kappa$ and $\xi$. 


While classical field theory is typically over spacetime, poly-KvN mechanics motivates classical fields over phase space. 
The 0th quantized poly-KvN algebra is therefore
\begin{eqnarray}
{[}\phi(x,p_x), \kappa(y,p_y)] &=& i\delta^{(4)}(x-y)\delta^{(4)}(p_x - p_y), \\ 
{[}\xi_\mu(x,p_x), \pi^\nu(y,p_y)] &=& i\delta_\mu^\nu \delta^{(4)}(x-y)\delta^{(4)}(p_x-p_y), \\ 
{[}\phi(x,p_x), \pi^\mu(y,p_y)] &=& 0, \\
{[}\xi_\mu(x,p_x), \kappa(y,p_y)] &=& 0,
\end{eqnarray}
where all other commutation relations vanish. 
This algebra contains fields that differ from standard classical field theory, as the Klein-Gordon scalar field typically is $\phi(x)$, not $\phi(x,p)$. 

Assuming $\phi(x)$ is a 1st-quantized field, the introduction of fields $\kappa(x)$ and $\xi_\sigma(x)$ over spacetime implies that the following covariant and canonical commutation relations should hold
\begin{eqnarray}
{[}\phi(x), \kappa(y)] &=& i\delta^4(x-y), \\ 
{[}\xi_\mu(x), \pi^\nu(y)] &=& i\delta_\mu^\nu \delta^4(x-y), \\ 
{[}\phi(x), \pi^\mu(y)] &=& 0, \\
{[}\xi_\mu(x), \kappa(y)] &=& 0. 
\label{KvNDDWalg1}
\end{eqnarray}
The units of $\kappa(x,p)$ must differ from $\kappa(x)$ if we assume that the units of $\phi(x)$ and $\phi(x,p)$ are the same. 
To understand second quantization of fields, the classical fields satisfying the first-quantized classical commutation relations are the most helpful. 

The dimensional analysis of fields over phase space differs from fields over spacetime. Consider a 0th quantized Klein-Gordon scalar field $\phi(x,p)$. If such a theory is to exhibit phase space symmetry with Born's reciprocity, then the Klein-Gordon scalar would be massless if we consider that $Spin(4,4)$ or $Spin(6,2)$ for relativistic phase space both contain the conformal group $Spin(4,2)$. The conformal group has been studied as a shadow of phase space \cite{Bars:2010xi} and conformal theories are massless. This ensures that the action principle can be considered without $\hbar$, since the kinetic term is void of $\hbar$, while the mass term is $-\frac{1}{2}\left(\frac{mc}{\hbar}\right)^2 \phi^2$. This gives $\phi(x,p)$ the dimensions of $\sqrt{\frac{m}{t}}$. The conjugate poly-momentum $\pi^\mu(x,p)$ has dimensions of $\frac{1}{l}\sqrt{\frac{m}{t}}$. The dimensional analysis of 0th and 1st quantized fields are shown in Table \eqref{dimensions}. 
\begin{table}
\begin{tabular}{c||c|c|c}
Field & 1st Quantized & Dimensions & Mass Dimensions\\ \hline
$\phi(x,p)$ & False & $\sqrt{\frac{m}{t}}$ & 1 \\ 
$\pi^\mu(x,p)$ & False & $\frac{1}{l}\sqrt{\frac{m}{t}}$ & 2 \\ 
$\kappa(x,p)$ & False & $\frac{1}{l^8}\left(\frac{t}{m}\right)^{\frac{9}{2}}$ & -1 \\ 
$\xi_\nu(x,p)$ & False & $\frac{1}{l^7}\left(\frac{t}{m}\right)^{\frac{9}{2}}$ & -2 \\ 
$\phi(x)$ & True & $\sqrt{\frac{m}{t}}$ & 1 \\ 
$\pi^\mu(x)$ & True & $\frac{1}{l}\sqrt{\frac{m}{t}}$ & 2 \\ 
$\kappa(x)$ & True & $\frac{1}{l^4}\sqrt{\frac{t}{m}}$ & 3\\ 
$\xi_\nu(x)$ & True & $\frac{1}{l^3}\sqrt{\frac{t}{m}}$ & 2 \\ 
\end{tabular}
\caption{The dimensional analysis of 0th and 1st quantized fields in KvNdDW mechanics are depicted above.}
\label{dimensions}
\end{table}
Assuming that the dimensions of $\phi(x,p)$ and $\phi(x)$ are the same, then the dimensions of $\int d^4p \kappa(x,p)$ are equal to $\kappa(x)$. First quantization of fields over phase space may involve integrating out energy-momentum, as
\begin{equation}
\int d^4p_y [\phi(x,p_x),\kappa(y,p_y)] = i\delta^4(x-y)\int d^4p_y \delta^4(p_x-p_y) = [\phi(x),\kappa(y)].
\end{equation}

In general, for the ``phase space field basis,'' the conjugate fields as operators for 0th quantized fields are given by
\begin{eqnarray}
\kappa(x,p_x) &=& -i \delta^{(4)}(x-y) \delta^{(4)}(p_x-p_y) \frac{\partial}{\partial \phi(y,p_y)}, \\ 
\xi_\mu(x,p_x) &=& i\delta^{(4)}(x-y) \delta^{(4)}(p_x-p_y) \frac{\partial}{\partial \pi^\mu(y,p_y)},
\end{eqnarray}
while for 1st quantized fields,
\begin{eqnarray}
\kappa(x) &=& -i \delta^{(4)}(x-y) \frac{\partial}{\partial \phi(y)}, \\ 
\xi_\mu(x) &=& i\delta^{(4)}(x-y) \frac{\partial}{\partial \pi^\mu(y)}.
\end{eqnarray}






\subsection{KvN quantization of De Donder-Weyl theory}
Next, quantum fields in the De Donder-Weyl formalism are discussed. Rather than referring to a particle's position and momentum, a quantum field $\Phi(x)$ replaces position, which admits a canonical poly-momentum $\Pi_\mu(x)$. 
In principle, one may consider a mixture of $\phi$ and $\xi^\sigma$ as well as $\pi^\mu$ and $\kappa$, but the mismatch in tensor rank implies the following quantization schemes may be considered,
\begin{eqnarray}
\Phi &=& a \phi + b^\mu \xi_\mu, \\
\Pi^\mu &=& c \pi^\mu + d^\mu \kappa.
\end{eqnarray}
The commutator of the second-quantized fields $\Phi(x)$ and $\Pi^\mu(y)$ for arbitrary coefficients is found from Eq.~\eqref{KvNDDWalg1}
\begin{equation}
[\Phi(x), \Pi^\mu(y) ] = \left( ad^\mu + b^\mu c\right) i \delta^{(4)}(x-y).
\end{equation}
The simplest scheme is $a  =1$, $d^\mu \propto \hat{\tau}^\mu$, $b^\mu = c= 0$. Next, we seek a solution for $d^\mu$ given $a=1$. If the dimensions of 1st and 2nd quantized fields $\phi(x)$ and $\Phi(x)$ are the same, this is analogous to the dimensions of $\hat{x}^\mu$ matching the dimensions of $\hat{X}^\mu$. However, $\kappa(x)$ and $\Pi^\mu(x)$ do not differ by dimensions of $\hbar$. The dimensions of $d^\mu$ must be the same as $\hbar c \tau^\mu$. One cannot set $\Pi^\mu(x) = \hbar \kappa(x)$ due to dimensional analysis and the poly-symplectic geometry, but $\Pi^\mu(x) = \hbar c\tau^\mu \kappa(x)$ is possible. 

The commutation relations of a field $\Phi$ and its conjugate poly-momentum $\Pi^{\mu}$ can be found by recognizing that $\Pi_D =\frac{\partial \mathcal{L}}{\partial \frac{\partial \Phi}{\partial t}} = \frac{1}{c} \Pi^{0}$ is Dirac's canonical momentum, where
\begin{equation}
[\Phi(x), \Pi_D(y)]_{x^{0}=y^{0}}=i \frac{\hbar}{c} \delta^{(3)}(x-y) . \label{Dirac}
\end{equation}
While this reference to $x^{0}$ and $y^{0}$ implies a specific frame, the canonical commutation relations in with the poly-momentum can utilize $\hat{\tau}^{\mu}$, giving
\begin{equation}
\left[\Phi(x), \Pi^{\mu}(y)\right]=i \hbar c \tau^{\mu} \delta^{(4)}(x-y)=i \hbar c|\tau| \hat{\tau}^{\mu} \delta^{(4)}(x-y),
\end{equation}
where $[\Phi(x), \Phi(y)]=\left[\Pi^{\mu}(x), \Pi^{\nu}(y)\right]=0$. These commutation relations are manifestly Lorentz invariant. Since $\tau^{\mu}$ is not meant to encode the time coordinate, but rather the direction of a time-like curve in relation to another frame, $|\tau|$ should never be zero. The magnitude relates to differences in time between spacetime foliations, which may motivate Planck's time as the interval to be used in Feynman's discrete path integral. The discrete calculus of time evolution has been discussed previously by one of the authors \cite{Kauffman2003,Kauffman2022}. The canonical commutation relations are recovered in the frame where $\hat{\tau}^{\mu}=(1,0,0,0)$,
\begin{eqnarray}
{\left[\Phi(x), \frac{1}{c}\Pi^{0}(y)\right] } &=& i \hbar |\tau| \delta^{(4)}(x-y)=i \hbar |\tau| \delta^{(3)}(x-y) \delta\left(t_{x}-t_{y}\right) \\
&=& i \hbar \delta^{(3)}(x-y) |\tau| \delta\left(c\left(t_{x}-t_{y}\right)\right)=i \hbar \delta^{(3)}(x-y) \delta\left(\left(t_{x}-t_{y}\right) /|\tau|\right).
\end{eqnarray}
Integrating over $t_y$ leads to $t_y = t_x$ and recovers Eq.~\eqref{Dirac}. In conclusion, the second-quantized De Donder-Weyl algebra leads to Dirac's canonical quantization when the zeroth component of the poly-momentum is considered and $\Phi(x)$ and $\Pi^0(y)$ are evaluated at the same time.


\end{document}